\documentclass[%
 reprint,
 superscriptaddress,
 amsmath,amssymb,
prb,
]{revtex4-2}
\usepackage{comment}
\usepackage{amsfonts}
\usepackage{graphicx}
\usepackage{physics}
\usepackage{color}
\usepackage{amsthm}
\usepackage{mathdots}
\usepackage{tikz}
\usepackage{tikz-cd}
\usepackage{relsize}
\usepackage{hyperref}
\usepackage{caption}
\usepackage{subcaption}
\usepackage{dutchcal}
\usepackage{algpseudocode,algorithm}
\usepackage{CJK}

\newcommand{\beq}{\begin{equation}\begin{aligned}}
\newcommand{\eeq}{\end{aligned}\end{equation}}

\usepackage{tikz}
\usepackage{tikz-cd}
\usetikzlibrary{decorations.pathmorphing}
\tikzset{snake it/.style={decorate, decoration=snake}}
\usetikzlibrary{calc}
\usepackage{tikz-3dplot} 
\tdplotsetmaincoords{70}{120}  
\tdplotsetrotatedcoords{+100}{0}{0} 
\usepackage{tikz}
\pgfdeclarelayer{bg}
\pgfsetlayers{bg,main}

\def\blue#1{{\color{black} #1}}
\def\red#1{{\color{black} #1}}
\def\white#1{{\color{white} #1}}

\usepackage[normalem]{ulem} 
\usepackage{soul}

\begin{document}

\begin{CJK*}{UTF8}{}
\title{Kennedy-Tasaki transformation and non-invertible symmetry in  lattice models beyond one dimension}
\author{Aswin Parayil Mana}
\affiliation{C. N. Yang Institute for Theoretical Physics, State University of New York at Stony Brook, New York 11794-3840, USA}
\affiliation{Department of Physics and Astronomy, State University of New York at Stony Brook, New York 11794-3840, USA}

\author{Yabo Li (\CJKfamily{gbsn}李雅博)}
\affiliation{C. N. Yang Institute for Theoretical Physics, State University of New York at Stony Brook, New York 11794-3840, USA}
\affiliation{Department of Physics and Astronomy, State University of New York at Stony Brook, New York 11794-3840, USA}

\author{Hiroki Sukeno (\CJKfamily{min}助野裕紀)}
\affiliation{C. N. Yang Institute for Theoretical Physics, State University of New York at Stony Brook, New York 11794-3840, USA}
\affiliation{Department of Physics and Astronomy, State University of New York at Stony Brook, New York 11794-3840, USA}

\author{Tzu-Chieh Wei (\CJKfamily{bsmi}魏子傑)}
\affiliation{C. N. Yang Institute for Theoretical Physics, State University of New York at Stony Brook, New York 11794-3840, USA}
\affiliation{Department of Physics and Astronomy, State University of New York at Stony Brook, New York 11794-3840, USA}

\date{\today}

\begin{abstract}
We give an explicit operator representation (via a sequential circuit and projection to symmetry subspaces) of Kramers-Wannier duality transformation in higher-dimensional subsystem symmetric models, generalizing the construction in the 1D  transverse-field Ising model. Using the Kramers-Wannier duality operator, we also construct the Kennedy-Tasaki transformation that maps subsystem symmetry-protected topological phases to spontaneous subsystem symmetry-breaking phases, where the symmetry group for the former is either $\mathbb{Z}_2\times\mathbb{Z}_2$ or $\mathbb{Z}_2$. This also generalizes the recently proposed picture of the one-dimensional Kennedy-Tasaki transformation as a composition of manipulations involving gauging and stacking symmetry-protected topological phases to higher dimensions.

\end{abstract}

\maketitle
\end{CJK*}

\section{Introduction}

Symmetry-protected topological phases (SPT phases)~\cite{guwen2009tensor,pollmann2010entanglement} have gained the attention of quantum information and condensed matter researchers for the last decade. 
SPT phases are the equivalence classes of gapped Hamiltonians with a given symmetry. 
We say two Hamiltonians are in the same phase if they can be connected by a path in the space of gapped Hamiltonians that respect the symmetry. 
There is also an equivalent definition in terms of states without mentioning Hamiltonians. These states are short-range entangled and possess trivial topological order. They were first discussed in the context of the spin-1 Haldane phase in the 1+1 dimension~\cite{haldane1983nonlinear}. 
Short-range entangled states in the same phase can be connected by a symmetric finite-depth quantum circuit. Conversely, no such circuit exists if they are in different SPT phases. The notion of SPT phases has been generalized to higher dimensions in bosonic systems classified by group cohomology~ \cite{chen2013symmetry,chen2012symmetry} and in fermionic SPTs by supergroup cohomology~\cite{gu2014symmetry} and more generally by cobordism~\cite{kapustin2014symmetry}. 
Furthermore, SPT phases protected by subsystem symmetries were explored in Refs.~\cite{you2018subsystem,devakul2018classification,raussendorf2019computationally}. 

Conventional spontaneous symmetry breaking (SSB)  phases, in contrast to SPT phases, can be described by local order parameters, with the famous example of the 2D classical Ising model~\cite{mccoy1973two,mccoy2009advanced}, as well as its corresponding 1D quantum Ising model in a transverse field. In these models, the so-called Kramers-Wannier (KW) duality~\cite{kramers1941statistics,onsager1944crystal,Kaufman1949,kogut1979introduction} maps the model to one on the dual lattice between high-temperature (or high-field, i.e., disordered) and low-temperature (or low-field, i.e., ordered) phases. There is a recent re-emergence of interest in the KW duality of the 1D transverse-field Ising model, which is regarded as a transformation on the same lattice, i.e., in the same Hilbert space, exchanging the Ising interaction and transverse-field terms.  In addition to the $Z_2$ spin-flip symmetry, at the critical point, an additional symmetry is hence given by the KW duality. Recently, an explicit expression for this symmetry action in terms of unitaries and projection was obtained by Refs.~\cite{ho2019efficient,seiberg2023majorana,chen2023sequential}. Due to the explicit projection onto the symmetric subspace, this is a non-invertible symmetry. Furthermore, this symmetry action squares to a lattice translation by one site times the projection~\cite{seiberg2023majorana}. Seiberg and Shao also show that this non-invertible symmetry comes from gauging the fermion parity of free Majorana fermions~\cite{seiberg2023majorana}. Majorana translation symmetry emerges as the non-invertible KW duality symmetry after gauging the fermion parity. This non-invertible symmetry fits with the non-invertible duality line in the Ising CFT~\cite{chang2019topological,frohlich2004kramers,frohlich2007duality,schutz1993duality}.

The Haldane phase, now recognized as an SPT phase, was found by Kennedy and Tasaki (KT) to relate via a non-local transformation to a spontaneous $\mathbb{Z}_2\times\mathbb{Z}_2$ symmetry breaking phase~\cite{kennedy1992hidden,kennedy1992hidden2}. Oshikawa explicitly constructed such a non-local transformation and generalized it to all integer spins~\cite{oshikawa1992hidden}. The overall picture that emerges is that the $\mathbb{Z}_2\times\mathbb{Z}_2$ SPT is mapped to two copies of the symmetry broken phase under the KT transformation~\cite{li2023non}. An explicit example given in Ref.~\cite{li2023non} is between the spin-1/2 1D cluster-state Hamiltonian, which represents a nontrivial $\mathbb{Z}_2\times\mathbb{Z}_2$ SPT phase, and two copies of the transverse-field Ising model, which represent $\mathbb{Z}_2$ SSB phases. Recently, there have also been generalizations of KT transformation to categorical symmetries~\cite{bhardwaj2023club} and in the context of LDPC codes \cite{rakovszky2023physics}.

From a different perspective, cluster states are resources for measurement-based quantum computation (MBQC)~\cite{raussendorf2001one,raussendorf2003measurement}, where performing single-qubit measurements
on cluster states leads to universal computation. In searching for order parameters to characterize quantum computational phases of matter for MBQC, Doherty and Bartlett~\cite{doherty2009identifying} considered a mapping that takes the 1D open-boundary cluster Hamiltonian to two copies of open-boundary transverse-field Ising models, which, in fact, realizes the same picture above of the KT transformation.
 They also mapped the 2D cluster state on a square lattice with a transverse field to two copies of the 2D transverse-field plaquette Ising model. Later, this transformation was generalized to 3D by You et al. in Ref.~\cite{you2018subsystem}, where they map the 3D cluster state to two copies of the 3D transverse-field cubic Ising model.  
 In these cases, we note that the nonlocal transformation involves mapping the Pauli $Z$ operators to an operator supported on one side of the light cone of Pauli $Z$. 
 We remark that there are a few ways to obtain the KT transformation, as will be discussed in section~\ref{sec:KTtwoandhigher}, and in fact, the exact mapping by Doherty and Bartlett can be derived.

 In this paper, we generalize the explicit operator construction of the KW duality symmetry to higher-dimensional hypercubic Ising models that have subsystem line-like symmetry. Such duality maps between the SSB and disordered phases.
We also discuss a fermionic dual to these models: Majorana hypercubic models that have subsystem fermion parity symmetry. We gauge the subsystem fermion parity to obtain the hypercubic Ising models. 
Connecting SPT to SSB phases, we provide an explicit operator representation of the KT transformation in one and higher dimensions. 
The KT transformation maps the $\mathbb{Z}_2\times\mathbb{Z}_2$-symmetric cluster-state model on the hypercubic bipartite lattice to two copies of hypercubic Ising models. We generalize the picture provided by Ref.~\cite{cao2022boson} for KT transformation to that between $\mathbb{Z}_2\times\mathbb{Z}_2$ subsystem  SPT (SSPT) to (two copies of) $\mathbb{Z}_2$ subsystem SSB (SSSB) phases in higher dimensions. We also discuss about $\mathbb{Z}_2$ subsystem symmetric model in two and higher dimensions and their corresponding $\mathbb{Z}_2$ subsystem SPT models. Specifically, we provide an explicit operator representation of the KW duality symmetry for the double hypercubic Ising model (DHCIM), and using this, we construct the KT transformation that maps between $\mathbb{Z}_2$ SSPT and one copy of DHCIM (which has SSSB).

The remaining structure of this paper is as follows. We give an overview of the results in Ref.~\cite{seiberg2023majorana} in section~\ref{sec:NIKT1D}, where they obtain an explicit expression for KW duality on a lattice. Using these results, we give an explicit expression for Kennedy-Tasaki transformation in one dimension mapping $\mathbb{Z}_2\times\mathbb{Z}_2$ SPT to two copies of the Ising model. We
also discuss the composition of KT and KW. In section~\ref{sec:KWtwoandhigher}, we discuss our results on the explicit KW duality operator for subsystem symmetric models in two and higher dimensions. Section~\ref{sec:KTtwoandhigher} deals with the KT transformation that maps the $\mathbb{Z}_2\times\mathbb{Z}_2$ SSPT to two copies of hypercubic Ising models that are in the SSSB phase. We mention the composition of KW and KT in two and higher dimensions in section~\ref{sec:Comptwoandhigher}. The discussion of KW and KT transformation for $\mathbb{Z}_2$ subsystem symmetric model in two and higher dimensions is presented in section~\ref{sec:z2subsymmetricmodel}. Finally, in section~\ref{sec:conc}, we conclude our results and give some future directions. Some discussions about the fermionic duals of the hypercubic Ising models are given in the appendix~\ref{sec:fermionicdual}. We discuss an alternate way to obtain the KT transformation in the appendix~\ref{sec:TST}. In appendix~\ref{sec:Z_Nhigherdimmodels}, we discuss $\mathbb{Z}_N$ generalization of hypercubic clock models and their non-invertible symmetries. In appendix~\ref{sec:Z_NgeneralKT}, we provide the $\mathbb{Z}_N$ generalization of KT transformation that maps $\mathbb{Z}_N\times\mathbb{Z}_N$ SPT phase to two copies of $\mathbb{Z}_N$ symmetry breaking phase in one dimension and discuss a KT transformation that maps $\mathbb{Z}_N\times \mathbb{Z}_N$ SSPT phase to two copies of $\mathbb{Z}_N$ subsystem symmetry breaking phase in two dimensions. We discuss a way to obtain the Hamiltonian of $\mathbb{Z}_2$ SSPT phase from the Hamiltonian of $\mathbb{Z}_2\times\mathbb{Z}_2$ SSPT by breaking the $\mathbb{Z}_2\times\mathbb{Z}_2$ symmetry to the diagonal subgroup in appendix~\ref{sec:symmetrybreaking}. In appendix~\ref{app:Measurement-based}, we describe how one can implement the KW and KT transformations in short-depth operations using gates and measurements.
\red{We discuss KT transformations between order parameters in Appendix~\ref{ref:order-param}.}

\section{Non-invertible symmetry and Kennedy-Tasaki transformation in one-dimensional lattice model}\label{sec:NIKT1D}
In this section, we review non-invertible symmetry and Kennedy-Tasaki transformation in one-dimensional lattice models. 
\subsection{Kramers-Wannier duality}
\subsubsection{$\mathbb{Z}_2$ symmetric model}\label{sec:KW-1d-Z2}

The two-dimensional classical Ising model possesses the famous Kramers-Wannier (KW) duality between high-temperature and low-temperature phases \cite{kramers1941statistics}. The duality maps an ordered phase to a disordered phase and vice versa. At the critical temperature, the KW duality is a self-duality, and that in turn determines the critical temperature. The two-dimensional classical model can be mapped to a one-dimensional quantum model by the quantum-classical correspondence \cite{kogut1979introduction}. Hence, one can discuss the KW duality in the context of a one-dimensional quantum lattice model, which is the transverse field Ising model (TFI). Consider the Hamiltonian of the TFI model on a ring with $L$ sites with a coupling parameter $\lambda$
\begin{align}
    \bm H_{\text{TFI}}=-\sum_{i=1}^L Z_iZ_{i+1}-\lambda \sum_{i=1}^L X_i.
\end{align}
Here, $X$, $Y$, and $Z$ are the Pauli operators. 
They are bosonic operators that anticommute with each other and square to identity. 
The model has a global $\mathbb{Z}_2$ symmetry operator $\eta\equiv\prod_{j=1}^L X_j$ and it commutes with the Hamiltonian $[\bm H_{\text{TFI}},\eta]=0$. 
This model at $\lambda=1$ also has a symmetry that interchanges the Ising and transverse field term. 
This is the quantum version of
the celebrated KW self-duality at the critical point for the classical model. 
Recently, an explicit expression for the KW duality symmetry for TFI was obtained by Ref.~\cite{ho2019efficient,seiberg2023majorana,chen2023sequential}. 
The KW duality which we denote by $\mathbf{D}$ can be decomposed into a unitary part and a projection onto the symmetric subspace where $\eta=1$. 
The unitary part that we denote by $\tilde{\mathbf{D}}$ is
an ordered product
of operators (specifically, a Clifford circuit) 
\begin{equation}
\label{eq:Dtilde}
\tilde{\mathbf{D}}\equiv\Big(\prod_{j=1}^{L-1} e^{i\frac{\pi}{4}X_j}e^{i\frac{\pi}{4}Z_jZ_{j+1}}\Big) e^{i\frac{\pi}{4}X_L},
\end{equation}
\begin{figure}[h!]
    \centering
    \includegraphics[scale=1]{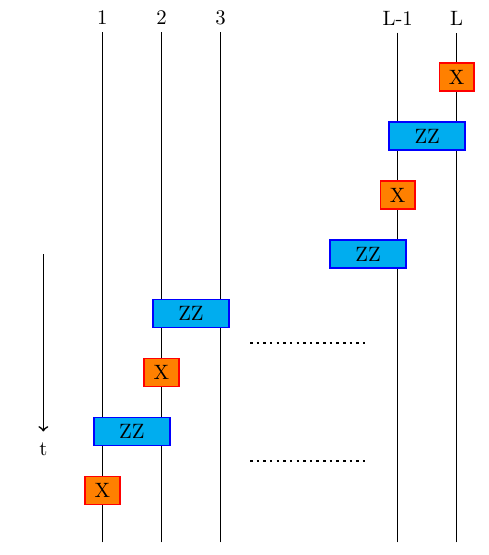}
    \caption{Quantum circuit representation of the operator $\tilde{\mathbf{D}}$ given in Eq.~\eqref{eq:Dtilde}. The time ordering of the circuit is from top to bottom. The top layer denotes the input sites $1$,$2$,...,$L$. The bottom layer represents the output.}
    \label{fig:1DKW}
\end{figure}
whereas the projector is defined by 
 \begin{align}
     \mathbf{P}\equiv \frac{(1+\eta)}{2}.
 \end{align}
The KW duality symmetry is the product $\tilde{\mathbf{D}}\mathbf{P}$:
\begin{align}
\mathbf{D} \equiv \tilde{\mathbf{D}} \mathbf{P} . \label{eq:1dKW}
\end{align}
It is a symmetry at $\lambda=1$ because it commutes with the Hamiltonian $[\bm H_{\text{TFI}},\mathbf{D}]=0$ as it exchanges the two terms in the Hamiltonian,
 \begin{equation}
\label{eq:DZ}
\mathbf{D} X_j = Z_j Z_{j+1} \mathbf{D}, \ \ X_{j+1} \mathbf{D} = \mathbf{D} Z_j Z_{j+1}.
\end{equation}
One can show easily that by using the unitary operator $\tilde{\mathbf{D}}$ it leads to $\tilde{\mathbf{D}} X_j \tilde{\mathbf{D}}^\dagger=Z_j Z_{j+1}$ for $j\ne L$,  $\tilde{\mathbf{D}} X_L \tilde{\mathbf{D}}^\dagger=Z_L Z_{1} \eta$, $\tilde{\mathbf{D}}^\dagger X_j \tilde{\mathbf{D}} =Z_{j-1} Z_{j}$ for $j\ne 1$, and $\tilde{\mathbf{D}}^\dagger X_1 \tilde{\mathbf{D}} =Z_{L} Z_{1}\eta$. The extra $\eta$ factor in the special boundary cases requires the introduction of the projector $(1+\eta)/2$ in $\mathbf{D}$ above, making the latter a non-invertible symmetry for the critical transverse-field Ising model.
Additionally, the action on $Z_j$ by $\tilde{\mathbf{D}}$ is: 
\begin{equation}
\tilde{\mathbf{D}} Z_j \tilde{\mathbf{D}}^\dagger =  Y_1 \prod_{k=2}^j X_k,
\label{eq:DZ1}
\end{equation}
and that by $\tilde{\mathbf{D}}^\dagger$ is: 
\begin{equation}
\tilde{\mathbf{D}} ^\dagger Z_j \tilde{\mathbf{D}} = -\Big(\prod_{k=j}^{L-1} X_k\Big) Y_L.
\end{equation}
From Eq.~\eqref{eq:DZ}, we easily see that $\mathbf{D}^2 X_j= X_{j+1} \mathbf{D}^2$ and $\mathbf{D}^2 Z_jZ_{j+1}= Z_{j+1}Z_{j+2} \mathbf{D}^2$ (with the periodic boundary condition $L+1\equiv 1$). \red{Moreover, it could be easily checked that $\mathbf{D}^{\dagger}\mathbf{D}=\textbf{P}$.} Hence, $\mathbf{D}^2$ is a product of lattice translation by one site, which we denote by $\red{\mathbf{T}}$, and the projection $\mathbf{P}$ onto the symmetric subspace,
\begin{align}
    \mathbf{D}^2{\red \propto}\, \red{\mathbf{T}}\mathbf{P}.
    \label{eq:TP1d}
\end{align}
\red{The proportionality factor is a phase and is fixed by the action of $\mathbf{D}^2$ on the product state $\ket{+}^{\otimes L}$.
\begin{align}
    \mathbf{D}^2=e^{\frac{2\pi i L}{4}}\mathbf{T}\mathbf{P}.
\end{align}
}
As a note, translation operation cannot be constructed in constant depth quantum circuit~\cite{gross2012index}. This is consistent with the fact that $\tilde{\mathbf{D}}$ is a linear-depth quantum circuit. \red{We note that the 
phase factor $(e^{2\pi i N/8})^2$ related to an anomaly in the corresponding fermionic theory~\cite{seiberg2023majorana}
, and it does not affect the transformation of operators}.

\subsubsection{$\mathbb{Z}_N$ symmetric model}
We generalize the discussion to the quantum clock model with an $N$-dimensional qudit degree of freedom at each site. 
We introduce generalized Pauli operators which obey $Z|a\rangle = \omega^a |a\rangle$, $X|a\rangle = |a+1\text{ mod }N\rangle$ with $\omega = e^{\frac{2\pi i}{N}}$ so that $X^\dagger = X^{-1}$,$Z^\dagger = Z^{-1}$, $ZX=\omega XZ$, $Z^\dagger X = \omega^{-1} X Z^\dagger$, etc. 
Let the model be defined on a ring with $L$ sites. 
The Hamiltonian of the model is given by
\begin{widetext}
 \begin{align}
     H=-J\left(\sum_i \left(Z_i^{\dagger}Z_{i+1}+Z_iZ_{i+1}^{\dagger}+ \lambda (X_i+X_i^{\dagger})\right)\right),
 \end{align}
\end{widetext}
 with periodic identifications $Z_{i+L}\equiv Z_i$ and $X_{i+L}\equiv X_i$. 
This model has a global symmetry generated by $\eta=\prod_i X_i$. 
Note that $\eta^2$,..., $\eta^{N-1}$ are also global symmetries. 
The model is known to have a critical point describing phase transition at $\lambda=1$ for $N=2,3,4$ ~\cite{jose1977renormalization,ortiz2012dualities,chen2017phase}. 
At $\lambda=1$, there is a symmetry under the Kramers-Wannier transformation that 
 interchanges the interaction term and the transverse-field term. 
This self-dual point is described by the $\frac{SU(2)_N}{U(1)}$ coset conformal field theory with central charge $c=\frac{2(N-1)}{N+2}$~\cite{zamolodchikov1985nonlocal,fateev1991integrable} for $N=2,3,4$. (We comment that for $N>4$, the central charge is $c=1$~\cite{li2015criticality,li2020critical} and there is, in fact, an intermediate critical phase instead of a single transition point.) Let us denote the Kramers-Wannier duality symmetry for this model as $\mathbf{D}_{(N)}$. We define
\begin{widetext}
  \begin{align}
     \begin{split}
         \mathbf{D}_{(N)}&\equiv\mathbf{P}_{(N)}\tilde{\mathbf{D}}_{(N)}\mathbf{P}_{(N)}\equiv\mathbf{P}_{(N)}\prod_{j=1}^{L-1}\left[\left(\sum_{n=0}^{N-1} c(n)X_j^n\right)\left(\sum_{n=0}^{N-1} \Tilde{c}(n)(Z_jZ_{j+1}^{\red{\dagger}})^n\right)\right]\times\left(\sum_{n=0}^{N-1}c(n)X_L^n\right)\mathbf{P}_{(N)} ,
     \end{split}
     \label{eq:D_N0}
 \end{align}
 \end{widetext}
 where the projection operator is onto the $\mathbf{Z}_{N}$ symmetric state given by
 \begin{align}
 \mathbf{P}_{(N)}&\equiv\left(\frac{1+\eta+\eta^2+...+\eta^{N-1}}{N}\right) , 
 \end{align}
and the coefficients in Eq.~\eqref{eq:D_N0} are given by 
     \begin{align}
     \Tilde{c}(n)=\frac{\omega^{\red{-}\frac{n(N+n)}{2}}}{\red{\sqrt{N}}},\quad
     c(n)=\frac{\omega^{\red{}\frac{n(N-n)}{2}}}{\red{\sqrt{N}}}.
 \end{align}
 The operator $\mathbf{D}^{(N)}$ satisfies the following properties:
 \begin{subequations}
    \begin{align}
     \mathbf{D}_{(N)}X_j=Z_jZ_{j+1}^{\red{\dagger}} \mathbf{D}_{(N)} , \\
     \mathbf{D}_{(N)}X_j^{\dagger}=Z_j^{\red{\dagger}}Z_{j+1}\mathbf{D}_{(N)} , \\
     \mathbf{D}_{(N)}Z_jZ_{j+1}^{\red{\dagger}}=X_{j+1}\mathbf{D}_{(N)} , \\
\mathbf{D}_{(N)}Z_j^{\red{\dagger}}Z_{j+1}=X_{j+1}^{\dagger}\mathbf{D}_{(N)} . 
\end{align} 
\label{eq:D_N}
\end{subequations}
Moreover, $\mathbf{D}_{(N)}$ commutes with the Hamiltonian at $\lambda=1$, \red{$\mathbf{D}_{(N)}^{\dagger}\mathbf{D}_{(N)}=\mathbf{D}_{(N)}\mathbf{D}_{(N)}^{\dagger}=\mathbf{P}_{(N)}$} and 
\begin{align}
    \mathbf{D}_{(N)}^2= \red{e^{ 2i\phi(N,L)}}\red{\mathbf{T}_{clock}}\mathbf{P}_{(N)}\, .
\end{align}
 Here $\red{\mathbf{T}_{clock}}$ is the translation on the lattice by one site \red{and $e^{i\phi(N,L)}$ is a phase factor. For small values of $N$, we give the values of phase $e^{i\phi(N,L)}$ in Table.~\ref{tab:phi(N,L)values}.}   
\red{\begin{table}
    \centering
\begin{tabular}{|c|c|c|c|c|}
\hline
   $N$ & 2 & 3 & 4 & 5\\
\hline
    $e^{i\phi(N,L)}$ & $e^{\frac{2\pi i L}{8}}$ & $e^{-\frac{2\pi i L }{4}}$ & $e^{-\frac{2\pi i L}{8}}$ & 1\\
   \hline
\end{tabular}
\caption{\red{The values of phase $e^{i\phi(N,L)}$ for various choices of $N$.}}
    \label{tab:phi(N,L)values}
\end{table}}
For completeness, we note that the action of $\tilde{\mathbf{D}}_{(N)}$ (the unitary part of $\mathbf{D}_{(N)}$) on a single $Z$ operator is
\begin{align}
     \tilde{\mathbf{D}}_{(N)}Z_i=\omega^{\red{\frac{(N^2+1)}{2}}}Z_1\prod_{j=1}^iX_j^{\red{\dagger}} \tilde{\mathbf{D}}_{(N)},
\end{align}
which reduces to Eq.~\eqref{eq:DZ1} for $N=2$.
\subsection{Kennedy-Tasaki transformation}
Kennedy and Tasaki constructed a non-local transformation that maps a $\mathbb{Z}_2\times\mathbb{Z}_2$ symmetry-protected topological phase to a symmetry breaking phase in Ref.~\cite{kennedy1992hidden,kennedy1992hidden2}. 
They constructed the transformation in the context of $S=1$ spin chains.
However, a simple compact expression valid for any integer spin was found by Ref.~\cite{oshikawa1992hidden}. 

Here, we construct an explicit expression for the Kennedy-Tasaki transformation using the Kramers-Wannier duality symmetry operator $\mathbf{D}$. 
Let us consider a spin chain with $L=2M$ sites with $\mathbb{Z}_2\times\mathbb{Z}_2$ symmetry on the odd and even sublattices. 
The Hamiltonian for this spin chain is 
\begin{align}
   \bm H_{\text{cluster}}=-\sum_{i=1}^{2M}Z_{i-1}X_iZ_{i+1}-\lambda\sum_{i=1}^{2M}X_i
   \label{eq:Hcluster}.
\end{align}
This is the cluster Hamiltonian with a transverse field in one dimension. 
We assume the periodic boundary condition and hence $Z_{2M+1}= Z_{1}$ and also take $Z_0\equiv Z_{2M}$. 
We define the symmetry generator $\eta_{even}\equiv\prod_{k=1}^MX_{2k}$ and $\eta_{odd}\equiv\prod_{k=1}^MX_{2k-1}$ on the even and odd sublattices respectively. For both the even and odd sublattices we introduce the KW duality operators
\begin{widetext}
\begin{subequations}
\begin{align}
    &\mathbf{D}_{even}\equiv\Big(\prod_{k=1}^{M-1} e^{i\frac{\pi}{4}X_{2k}}e^{i\frac{\pi}{4}Z_{2k}Z_{2k+2}}\Big) e^{i\frac{\pi}{4}X_{2M}}\frac{(1+\eta_{even})}{2},\\
    &\mathbf{D}_{odd}\equiv\Big(\prod_{k=1}^{M-1} e^{i\frac{\pi}{4}X_{2k-1}}e^{i\frac{\pi}{4}Z_{2k-1}Z_{2k+1}}\Big) e^{i\frac{\pi}{4}X_{2M-1}}\frac{(1+\eta_{odd})}{2}.
\end{align}
\end{subequations}
\end{widetext}
\normalsize
Let us define the cluster entangler $T \equiv \prod_j CZ_{j,j+1}$, where $CZ$ is the controlled-Z gate $CZ_{j,j+1}\equiv e^{i\frac{\pi}{4}(1-Z_j)(1-Z_{j+1})}$. Note that the cluster entangler is equivalent to stacking an SPT phase.
With these ingredients, we define the Kennedy-Tasaki transformation as
\beq
    \text{KT}\equiv \mathbf{D}_{odd}^{\dagger}\mathbf{D}_{even}^{\dagger}T\mathbf{D}_{even}\mathbf{D}_{odd} . 
    \label{eq:1dKT}
\eeq

\red{Ref}.~\cite{li2023non} showed that the
KT transformation in the continuum is equivalent to a sequence of operations:  $TST$ or $STS$, where $S$ is gauging the global symmetry and $T$ is stacking a $\mathbb{Z}_2\times\mathbb{Z}_2$ SPT.
Our definition here uses the picture of $S^{\dagger}TS$ that is similar to the picture of $STS$ up to a translation in its transformation.   \red{Since we use a quantum circuit and projection to implement KW transformation, in our definition of $S$, we have a lattice translation by one unit. Therefore, $S^{\dagger}$ is not equal to $S$ on the lattice.} Note that both $S^{\dagger}TS$ and $STS$ are the same in the continuum since in the continuum translation by one unit goes to identity operation and  $S^{\dagger}$=$S$.

Our calculation verifies that $S^{\dagger}TS$ gives rise to the KT transformation.   
Explicitly, the action of our KT transformation is given by
\begin{subequations}
    \begin{align}
    \text{KT}\,X_i&=X_i\,\text{KT} , \\
    \text{KT}\,Z_{i-1}X_iZ_{i+1}&=Z_{i-1}Z_{i+1}\,\text{KT} . 
\end{align}
\label{eq:KTtrans1D}
\end{subequations}
Hence, KT maps the cluster Hamiltonian to two (decoupled) copies of Ising models in the two sublattices. On a single $Z$ operator, the action of $\text{KT}$ is given by 
\begin{subequations}
\begin{align}
    \text{KT}\,Z_{2i+1}&=Z_{2i+1}\prod_{k=i+1}^{M-1}X_{2k}\,\text{KT}', \\ 
    \text{KT}\,Z_{2i}&=\prod_{k=1}^{i}X_{2k-1}Z_{2i}\,\text{KT}'' ,
\end{align}
\end{subequations}
where $\text{KT}'$ and $\text{KT}''$ are defined as
\begin{subequations}
\begin{align}
    \text{KT}'\equiv \mathbf{D}_{odd}^{'\dagger}\mathbf{D}_{even}^{\dagger}T\mathbf{D}_{even}\mathbf{D}_{odd}',\\
    \text{KT}''\equiv \mathbf{D}_{odd}^{\dagger}\mathbf{D}_{even}^{'\dagger}T\mathbf{D}_{even}'\mathbf{D}_{odd},
\end{align}
\end{subequations}
with 
\begin{subequations}
\begin{align}
\begin{split}
    \mathbf{D}_{odd}'&\equiv\Big(\prod_{k=1}^{M-1} e^{i\frac{\pi}{4}X_{2k-1}}e^{i\frac{\pi}{4}Z_{2k-1}Z_{2k+1}}\Big)\\
    &\qquad\times e^{i\frac{\pi}{4}X_{2M-1}}\frac{(1-\eta_{odd})}{2},
    \end{split}\\
    \begin{split}
     \mathbf{D}_{even}'&\equiv\Big(\prod_{k=1}^{M-1} e^{i\frac{\pi}{4}X_{2k}}e^{i\frac{\pi}{4}Z_{2k}Z_{2k+2}}\Big)\\
    &\qquad\times e^{i\frac{\pi}{4}X_{2M}}\frac{(1-\eta_{even})}{2}.
    \end{split}
\end{align}
\label{eq:DoddDeven}
\end{subequations}
\normalsize
Similarly,
\begin{subequations}
\begin{align}
    \text{KT}^{\dagger}Z_{2i+1}=Z_{2i+1}\prod_{k=i+1}^{M\red{-1}}X_{2k}\,(\text{KT}')^{\dagger},\\
    \text{KT}^{\dagger}Z_{2i}=\prod_{k=1}^{i}X_{2k\red{-}1}Z_{2i}\,(\text{KT}'')^{\dagger}
    .
\end{align}
\end{subequations}
Composition of $\text{KT}$ with itself gives
\begin{align}
    \text{KT}^2=\frac{(1+\eta_{even})}{2}\frac{(1+\eta_{odd})}{2},
\end{align}
 The projection factor implies that $\text{KT}$ is non-invertible. 

As a remark, we can also use the following definition to implement the KT transformation,
\beq
    \text{KT}\equiv \mathbf{D}_{odd}\mathbf{D}_{even}T\mathbf{D}_{even}\mathbf{D}_{odd},
    \label{eq:1dKT-a}
\eeq
which differs from Eq.~\eqref{eq:1dKT} by the last part (without the Hermitian conjugation), and this will lead to additional translation.
Furthermore, we have also verified that the alternative definition corresponding to $TST$ also gives a valid KT transformation. We refer the readers to appendix~\ref{sec:TST}.

\subsection{Composition of operators}\label{subsec:Compoperators}
So far, we discussed the $\mathbf{D}$ operator, which implements the Kramers-Wannier transformation, and $\text{KT}$, which implements the Kennedy-Tasaki transformation in one dimension. 
Here, we consider the composition of operators
\begin{align}
    \begin{split}
        \mathbf{D}_{SPT}&\equiv\text{KT}^{\dagger}\red{\mathbf{D}_{even}\mathbf{D}_{odd}}\,\text{KT}\\
        &=\mathbf{D}_{odd}^{\dagger}\mathbf{D}_{even}^{\dagger}\,T\,\mathbf{D}_{even}\mathbf{D}_{odd}\,T\,\mathbf{D}_{even}\mathbf{D}_{odd}
    \end{split}
    \label{eq:DSPT1d}
\end{align}
where
\begin{figure*}
    \centering
\includegraphics[scale=1]{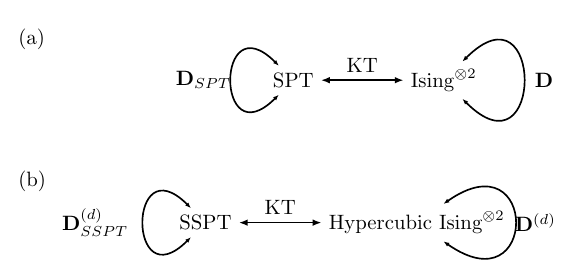}
\caption{(a) KT transformation from SPT to SSB; (b) KT transformation from SSPT to SSSB.}
\label{fig:composition}
\end{figure*}
this composition is schematically summarized in Fig.~\ref{fig:composition}(a).
Its action on the terms in the $\mathbb{Z}_2\times\mathbb{Z}_2$ cluster state Hamiltonian with the transverse field in Eq.~\eqref{eq:Hcluster} is 
\begin{subequations}
\begin{align}
    \mathbf{D}_{SPT}X_i=Z_iX_{i+1}Z_{i+2}\mathbf{D}_{SPT} ,\\
    \mathbf{D}_{SPT}Z_{i-1}X_iZ_{i+1}=X_{i+1}\mathbf{D}_{SPT}.
\end{align}
\label{eq:DSPT1transformation}
\end{subequations}
On single $Z$ operators, $\red{\mathbf{D}}_{SPT}$ act as
\begin{subequations}
\begin{align}
\mathbf{D}_{SPT}Z_{2i+1}&=\red{-}Z_1Z_{2i+2}Y_{2M}\mathbf{D}_{SPT}' , \\
    \mathbf{D}_{SPT}Z_{2i}&=\red{-}Y_1Z_2Z_{2i+1}\mathbf{D}_{SPT}'' ,
\end{align}
\end{subequations}
where 
\begin{subequations}
\begin{align}
    \mathbf{D}_{SPT}' \equiv\mathbf{D}_{odd}^{'\dagger}\mathbf{D}_{even}^{\dagger}\,T\,\mathbf{D}_{even}\mathbf{D}_{odd}'\,T\,\mathbf{D}_{even}\mathbf{D}_{odd}'\\
    \mathbf{D}_{SPT}''\equiv\mathbf{D}_{odd}^{\dagger}\mathbf{D}_{even}^{'\dagger}\,T\,\mathbf{D}_{even}'\mathbf{D}_{odd}\,T\,\mathbf{D}_{even}'\mathbf{D}_{odd}
\end{align}
\end{subequations}
with $\mathbf{D}_{odd}'$ and $\mathbf{D}_{even}'$ given in Eq.~\eqref{eq:DoddDeven}. 
We remark that $\mathbf{D}_{SPT}$ squares to the projection and translation by two sites
\blue{--- or equivalently, by one site in each sublattice.} 
Similar duality transformation \red{as in Eq.~\eqref{eq:DSPT1transformation}} can be achieved by $U_{CZ}\equiv \prod_{j} CZ_{j,j+1}$\red{, which is the cluster entangler $T$ defined above}, although $(U_{CZ})^2 =1$ and there is no translation.

\section{Non-invertible symmetry in lattice models beyond one dimension}\label{sec:KWtwoandhigher}

Here, we provide a generalization of non-invertible symmetry in higher dimensions. 
We will use inputs from the one-dimensional non-invertible symmetry construction for this generalization. The models we consider for the generalization of non-invertible symmetry in higher dimensions possess subsystem symmetries compared to the canonical example of a non-invertible symmetric system in one dimension (transverse-field Ising model) that possesses a 0-form symmetry.

\subsection{Two dimensions}
As a first generalization, we will look at two dimensions.
Recently, Ref.~\cite{cao2023subsystem} provided a construction of a subsystem non-invertible symmetry operator that maps between self-dual models on different lattices. Our KW construction here builds upon the 1D result by Seiberg and Shao~\cite{seiberg2023majorana}, which we have already reviewed in the section.~\ref{sec:KW-1d-Z2}; we map self-dual models on the same lattice.

In two dimensions, we consider the transverse-field plaquette Ising model, which is defined on a two-dimensional square lattice. 
Let us denote the coordinates of the square lattice by a pair $(i,j)$ where $i$ denotes the x-coordinate, and $j$ denotes the y-coordinate. 
The Hamiltonian for this model is given by
\begin{equation}   
\bm H_{\text{TFPI}}=-\sum_{i,j} (\,Z_{i,j} Z_{i+1,j} Z_{i+1,j+1} Z_{i,j+1}+ \lambda\, X_{i,j}).
\label{eq:HTFPI}
\end{equation}
This model has subsystem symmetries along horizontal and vertical directions, i.e., operators of the form $\eta^x_j\equiv\prod_{i}X_{i,j}$ and $\eta^y_i\equiv\prod_j X_{i,j}$ commute with the Hamiltonian. If we put the square lattice on a torus, with $L_x$ unit cells along $x$ direction and $L_y$ unit cells along $y$ direction, then there are $L_x+L_y$ symmetry generators. However, there is one constraint among them: $\prod_{j}\eta^x_j=\prod_i\eta^y_i$. Hence, we have $L_x+L_y-1$ independent symmetry generators. Additionally, at $\lambda=1$, there is an extra symmetry that exchanges the plaquette term and the transverse field. 
This is the generalization of the famous KW duality to the transverse-field plaquette Ising model. 
Here, we write down an expression for the KW duality operator on the square lattice,  generalizing the 1D construction in Ref.~\cite{seiberg2023majorana}. 

Let us now define the following operators
\begin{subequations}
\begin{align}
\tilde{\mathbf{D}}_x&\equiv\prod_{j=1}^{L_y}\left(\left(\prod_{i=1}^{L_x-1}e^{i\frac{\pi}{4}X_{i,j}}e^{i\frac{\pi}{4}Z_{i,j}Z_{i+1j}}\right)\,e^{i\frac{\pi}{4}X_{L_x,j}}\right), \label{Dx-def}\\
\tilde{\mathbf{D}}_y&\equiv\prod_{i=1}^{L_x}\left(\left(\prod_{j=1}^{L_y-1}e^{i\frac{\pi}{4}X_{i,j}}e^{i\frac{\pi}{4}Z_{i,j}Z_{i,j+1}}\right)e^{i\frac{\pi}{4}X_{i,L_y}}\right),\label{Dy-def}\\
\mathbf{P}^{\red{(2)}}&\equiv\prod_{j=1}^{L_y}\frac{(1+\eta^x_j)}{2}\prod_{i=1}^{L_x}\frac{(1+\eta^y_i)}{2}.
\end{align}
\label{Ddef}
\end{subequations}
As it can be seen, $\tilde{\mathbf{D}}_x$ is composed of rows of the one-dimensional version of $\tilde{\mathbf{D}}$ in Eq.~\eqref{eq:Dtilde}. Similarly, $\tilde{\mathbf{D}}_y$ is composed of columns of the one-dimensional version of $\tilde{\mathbf{D}}$ in Eq.~\eqref{eq:Dtilde}. $\mathbf{P}^{\red{(2)}}$ is a projector onto the subsystem symmetric subspace along both x and y directions. Now we define the following unitary action,
\begin{align}
    \Tilde{\mathbf{D}}^{(2)}\equiv\Tilde{\mathbf{D}}_x\red{\mathbf{H}^{\otimes(2)}}\Tilde{\mathbf{D}}_y ,  \label{eq:2DDsym}
\end{align}
where $\red{\mathbf{H}^{\otimes(2)}}$ represents the action of the Hadamard transformation on all sites (which transforms between $Z_{i,j}$ and $X_{i,j}$). \red{We \blue{comment} that the particular ordering of operators we \blue{set} in defining Eq.~\eqref{eq:2DDsym} is our choice. Another possible definition is $\Tilde{\mathbf{D}}^{(2)}\equiv\Tilde{\mathbf{D}}_y\red{\mathbf{H}^{\otimes(2)}}\Tilde{\mathbf{D}}_x$. We note that the two definitions are related by swapping the $x$ and $y$ coordinates. Their action on Pauli operators $X_{i,j}$ and $Z_{i,j}$ operators are related by swapping $(i,j)\longleftrightarrow(j,i)$.}

For $i \neq L_x$ and $j\neq L_y$, $\Tilde{\mathbf{D}}^{(2)}$ exchanges the two terms in the Hamiltonian as
\begin{subequations}
\begin{align}
    \Tilde{\mathbf{D}}^{(2)} X_{i,j}=Z_{i,j}Z_{i,j+1}Z_{i+1,j}Z_{i+1,j+1}\Tilde{\mathbf{D}}^{(2)},\\ \Tilde{\mathbf{D}}^{(2)}Z_{i,j}Z_{i,j+1}Z_{i+1,j}Z_{i+1,j+1}= X_{i+1,j+1}\Tilde{\mathbf{D}}^{(2)} .
\end{align}
\end{subequations}
Near the boundary, the transformation picks up subsystem symmetric factors:
\begin{subequations}
\begin{align}
    \Tilde{\mathbf{D}}^{(2)}X_{L_x,j}&=Z_{L_x,j}Z_{1,j}Z_{L_x,j+1}Z_{1,j+1}\eta^x_j\eta^x_{j+1}\Tilde{\mathbf{D}}^{(2)} , \\
    \Tilde{\mathbf{D}}^{(2)} X_{i,L_y}&=Z_{i,L_y}Z_{i+1,L_y}Z_{i,1}Z_{i+1,1}\Tilde{\mathbf{D}}^{(2)}\eta_i^y , \\
    \Tilde{\mathbf{D}}^{(2)}X_{L_x,L_y}&=Z_{L_x,L_y}Z_{1,L_y}Z_{L_x,1}Z_{1,1}\eta^x_{L_{\red{y}}}\eta^x_1\Tilde{\mathbf{D}}^{(2)}\eta^y_{L_{\red{x}}} , 
    \end{align}
    \end{subequations}
    and similarly,
    \begin{subequations}
    \begin{align}
    \Tilde{\mathbf{D}}^{(2)}Z_{L_x,j}Z_{L_x,j+1}Z_{1,j}Z_{1,j+1}&=X_{1,j+1}\eta^x_{j+1}\Tilde{\mathbf{D}}^{(2)} , \\
    \Tilde{\mathbf{D}}^{(2)}Z_{i,L_y}Z_{i,1}Z_{i+1,L_y}Z_{i+1,1}&=X_{i\red{+1},1}\Tilde{\mathbf{D}}^{(2)}\eta^y_i\eta^y_{i+1} , \\
    \Tilde{\mathbf{D}}^{(2)}Z_{L_x,L_y}Z_{L_x,1}Z_{1,L_y}Z_{1,1}&=X_{1,1}\eta^x_1
\Tilde{\mathbf{D}}^{(2)}\eta^y_1\eta^y_{L_x} . 
\end{align}
\end{subequations}
We can absorb the subsystem symmetric factors by introducing projection and defining the following  non-invertible operator 
\begin{align}
    \mathbf{D}^{(2)}\equiv\mathbf{P}^{\red{(2)}} \Tilde{\mathbf{D}}^{(2)} \mathbf{P}^{\red{(2)}}.
    \label{eq:D^2}
\end{align}
The operator $\mathbf{D}^{(2)}$ interchanges the plaquette term and the transverse field term, and we identify it as the KW duality transformation. In particular,
\begin{subequations}
\begin{align}
    \mathbf{D}^{(2)}X_{i,j}=Z_{i,j}Z_{i+1,j}Z_{i,j+1}Z_{i+1,j+1}\mathbf{D}^{(2)},\\
    \mathbf{D}^{(2)}Z_{i,j}Z_{i+1,j}Z_{i,j+1}Z_{i+1,j+1}=X_{i+1,j+1}\mathbf{D}^{(2)},
\end{align}
\end{subequations}
and hence $\mathbf{D}^{(2)}$ is a symmetry of the Hamiltonian Eq.~\eqref{eq:HTFPI} at $\lambda=1$, i.e., $ [\bm H_{\text{TFPI}}\rvert_{\lambda=1},\mathbf{D}^{(2)}]=0$. Action of $\mathbf{D}^{(2)}$ on a single $Z_{i,j}$ operator is
\begin{widetext}
\begin{align}
    \begin{split}
    \mathbf{D}^{(2)}Z_{i,j}&=(-i)Z_{i,1}Z_{i+1,1}Y_{1,1}\prod_{k=2}^iX_{k,1}\times \prod_{m=2}^j\left(Y_{1,m}\prod_{l=2}^iX_{l,m}\right)\mathbf{D}^{(2)'} ,
    \end{split}
\end{align}
where
\begin{align}
    \mathbf{D}^{(2)'}\equiv\tilde{\mathbf{P}}^{\red{(2)}'}\tilde{\mathbf{D}}^{(2)}\tilde{\mathbf{P}}^{\red{(2)}},
\end{align}
with
\begin{subequations}
\begin{align}
    \begin{split}
        \mathbf{\tilde{P}}^{\red{(2)}'}&\equiv\prod_{l=1}^j\frac{\left(1-\eta^x_l\right)}{2}\prod_{l=j+1}^{L_y}\frac{\left(1+\eta^x_l\right)}{2}\prod_{k\neq 1,i,i+1}\frac{\left(1+\eta^y_k\right)}{2}\times\frac{(1-\eta^y_i)}{2}\frac{(1-\eta^y_{i+1})}{2}\frac{(1+(-1)^j\eta^y_1)}{2},
    \end{split}\\
    \mathbf{\tilde{P}}^{\red{(2)}}&\equiv\prod_{l\neq j}\frac{(1+\eta^x_l)}{2}\prod_{k\neq i}\frac{(1+\eta^y_k)}{2}\frac{(1-\eta^x_j)}{2}\frac{(1-\eta^y_i)}{2},
\end{align}
\end{subequations}
and 
\begin{align}
    \begin{split}
        \mathbf{D}^{(2)\dagger}Z_{i,j}&=(-i)(-1)^{L_x-i+1}\prod_{k=i}^{L_x-1}\left(\left(\prod_{l=j}^{L_y-1}X_{k,l}\right)Y_{k,L_y}\right)\left(\prod_{l=j}^{L_y-1}X_{L_x,l}\right)Y_{L_x,L_y}Z_{L_x,j}Z_{L_x,j-1}\mathbf{\bar{P}}^{\red{(2)}'}\mathbf{\tilde{\mathbf{D}}}^{(2)\dagger}\mathbf{\bar{P}}^{\red{(2)}} ,
    \end{split}
\end{align} 
where
\begin{subequations}
\begin{align}
    \begin{split}
        \mathbf{\bar{P}}^{\red{(2)}'}&\equiv\prod_{k=1}^{i-1}\frac{(1+\eta_k^y)}{2}\prod_{k=i}^{L_x}\frac{(1-\eta_k^y)}{2}\prod_{l\neq j,j-1,L_y}\frac{(1+\eta_l^x)}{2}\frac{(1-\eta_j^x)}{2}\frac{(1-\eta_{j-1}^x)}{2}\frac{(1+(-1)^{L_x-i\red{+1}}\eta_{L_y}^x)}{2} ,
    \end{split}\\
    \mathbf{\bar{P}}^{\red{(2)}}&\equiv\prod_{k\neq i}\frac{(1+\eta_k^y)}{2}\prod_{l\neq j}\frac{(1+\eta_l^x)}{2}\frac{(1-\eta_{\red{i}}^y)}{2}\frac{(1-\eta_{\red{j}}^x)}{2}.
\end{align}
\end{subequations}
\end{widetext}
Moreover, after an appropriate choice \red{of normalization} for $\mathbf{D}^{(2)}$, $(\mathbf{D}^{(2)})^2$ is a translation $(i,j)\rightarrow (i+1,j+1)$ up to the projector:
\begin{align}
    (\mathbf{D}^{(2)})^2\propto\,\mathbf{T}_{(1,1)}\mathbf{P}^{\red{(2)}},
    \label{eq:TP2d}
\end{align}
where $\mathbf{T}_{(1,1)}$ is a translation along the diagonal. The projection $\mathbf{P}^{\red{(2)}}$ is onto the subsystem symmetric Hilbert space compared to the projection in Eq.~\eqref{eq:TP1d} that is onto a 0-form symmetric subspace. Note that the translation in the diagonal direction leads to traversing of all sites if the lengths $L_x$ and $L_y$ have no common factor, i.e., $\text{gcd}(L_x,L_y)=1$. 

We note that the KW transformation can also be implemented by unitary gates plus measurements in a short-depth operation, which maps the degrees of freedom to those on the dual lattice; see Appendix~\ref{app:Measurement-based}.
\subsection{Higher dimensions}

In higher dimensions, we consider the transverse-field hypercubic Ising model (TFHCI). 
This model is a natural generalization of the transverse field plaquette Ising model that we considered in two dimensions. 
It is defined on the hypercubic lattice with a hypercubic interaction of the vertices of a cube. 
In addition to the hypercubic Ising-like term, there is also a transverse field term. 

\begin{figure}
    \includegraphics[scale=1]{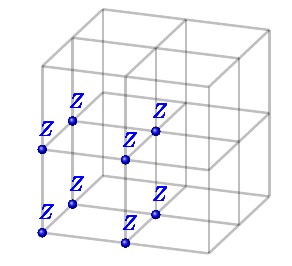}
    \caption{The Hamiltonian term $\prod_{v \in \partial c} Z_v$ in the transverse-field hypercube Ising model in three dimensions. }
    \label{fig:hypercube_Ising}
\end{figure}

Let us consider a hypercubic lattice in $d$ spatial dimensions with coordinates $x_1$,..., $x_d$. Let us denote the set of all vertices of the hypercubic lattice by $\Delta_v$ and the set of all hypercubes in the lattice by $\Delta_c$. We denote the boundary of a hypercubic cell $c$ by $\partial c$. The Hamiltonian for the TFHCI model is
\begin{align} \label{TFHCI}
    \bm H_{\text{TFHCI}}=-\sum_{c\in\Delta_c}\prod_{v\in\partial c}Z_{v}-\lambda\sum_{v\in\Delta_v}X_v.
\end{align}
See Fig.~\ref{fig:hypercube_Ising} for an illustration in three dimensions.
This model has subsystem line-like symmetries along all the coordinate directions. 
At $\lambda=1$, there is an additional symmetry that interchanges the hypercubic term and the transverse-field term. This is the generalization of KW duality to higher dimensions. We provide an explicit operator which achieves the KW duality similar to the two dimensions.
\begin{align}
    \mathbf{D}^{(d)}\equiv\mathbf{P}^{\red{(d)}} \tilde{\mathbf{D}}_d \red{\mathbf{H}^{\otimes(d)}} 
 \tilde{\mathbf{D}}_{d-1} \red{\mathbf{H}^{\otimes(d)}}\cdots\tilde{\mathbf{D}}_1 \mathbf{P}^{\red{(d)}},
 \label{eq:D^d}
\end{align}
 where each of the $\Tilde{\mathbf{D}}_{i}$ are string operators like Eq.~\eqref{eq:Dtilde} on a straight line along $i$-th direction\red{, $\red{\mathbf{H}^{\otimes(d)}}$ represents the tensor product of Hadamard operators on all the sites of $d$ dimensional lattice} and $\mathbf{P}^{\red{(d)}}$ is a projection onto subsystem straight line-like symmetries. The operator $\mathbf{D}^{(d)}$ is a symmetry of the TFHCI model as it commutes with the Hamiltonian, i.e., $[ \bm H_{\text{TFHCI}},\mathbf{D}^{(d)}]=0$. After an appropriate choice \red{of normalization} for $\mathbf{D}^{(d)}$, acting the operator twice will generate a diagonal unit shift in $d$ dimensions, i.e.,
 \begin{align}
     (\mathbf{D}^{(d)})^2\,\red{\propto}\,\mathbf{T}_{(1,...,1)}\mathbf{P}^{\red{(d)}}.
 \end{align}
\section{Kennedy-Tasaki transformation in lattice models beyond one dimension}\label{sec:KTtwoandhigher}
Here, we consider generalizations of Kennedy-Tasaki transformations to higher dimensions and derive them based on Fig.~\ref{fig:composition}. The KT transformation takes a higher-dimensional SPT phase to an SSB phase. Such an operator-mapping relation in 2D was first constructed in Ref.~\cite{doherty2009identifying} for the case of the open-boundary condition, which was then generalized to 3D in Ref.~\cite{you2018subsystem}. In this section, we study KT transformations that map $\mathbb{Z}_2\times \mathbb{Z}_2$ subsystem symmetric SPT phases to spontaneous subsystem symmetry breaking phases. Later in subsection~\ref{subsec:KTz_2SSPTtoSSSB}, we look at the KT transformations that map a $\mathbb{Z}_2$ SSPT phase to an SSSB phase, the latter of which breaks the $\mathbb{Z}_2$ subsystem symmetry.

\subsection{Two dimensions}
\label{sec:2dcluster}
\begin{figure*}[]
     \centering
    \begin{subfigure}[b]{0.6\columnwidth}
    \centering
    \includegraphics[width=\textwidth]{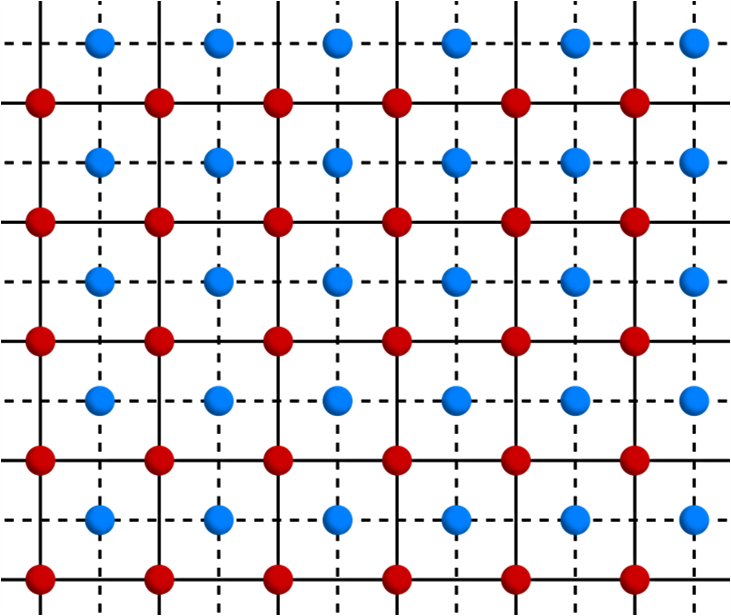}
    \caption{}
    \label{fig:2drblattice}
    \end{subfigure}
    \ \ \ \ \
    \begin{subfigure}[b]{0.6\columnwidth}
    \centering
    \includegraphics[width=\textwidth]{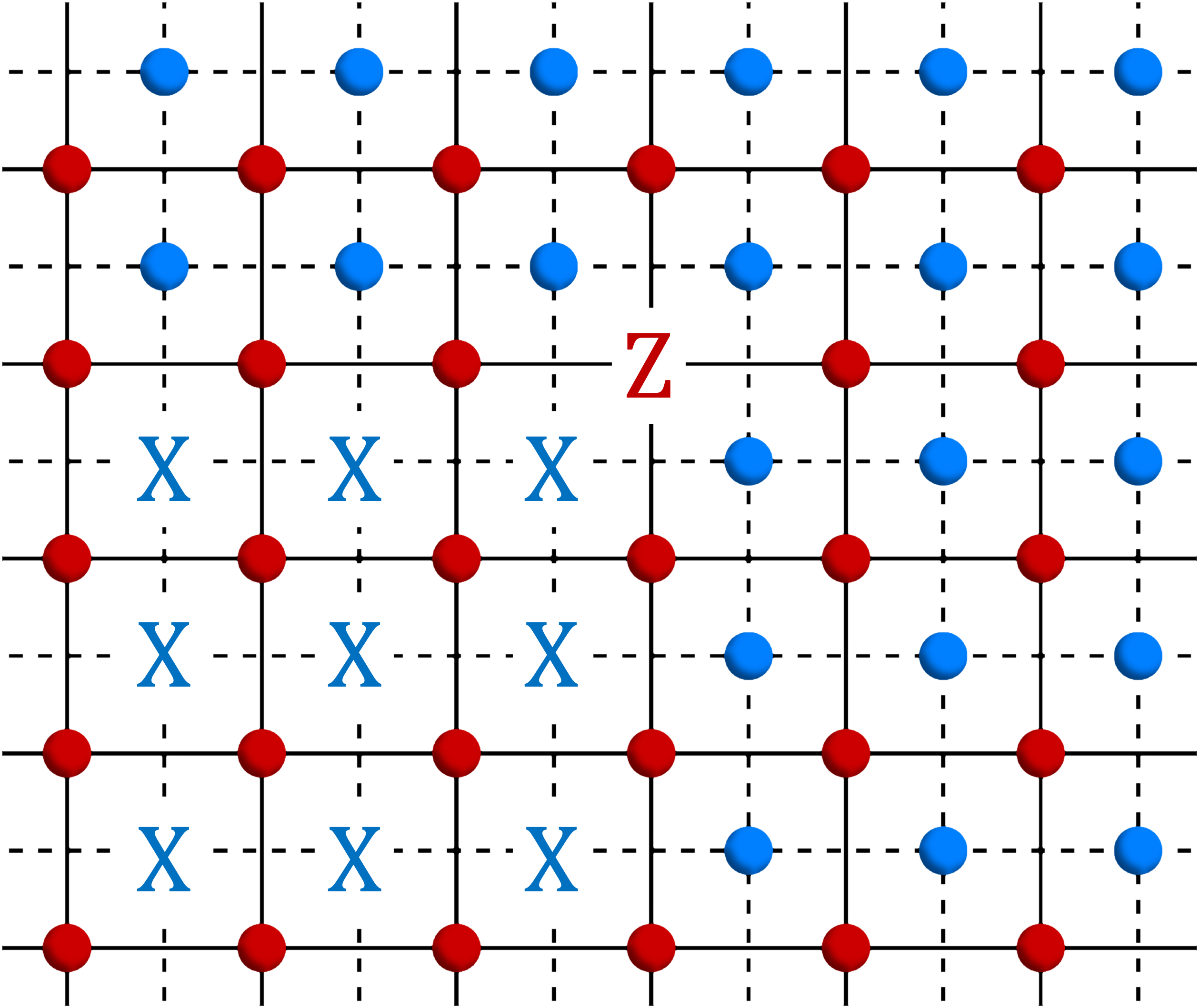}
    \caption{}
    \end{subfigure}
    \ \ \ \ \
    \begin{subfigure}[b]{0.6\columnwidth}
    \centering
    \includegraphics[width=\textwidth]{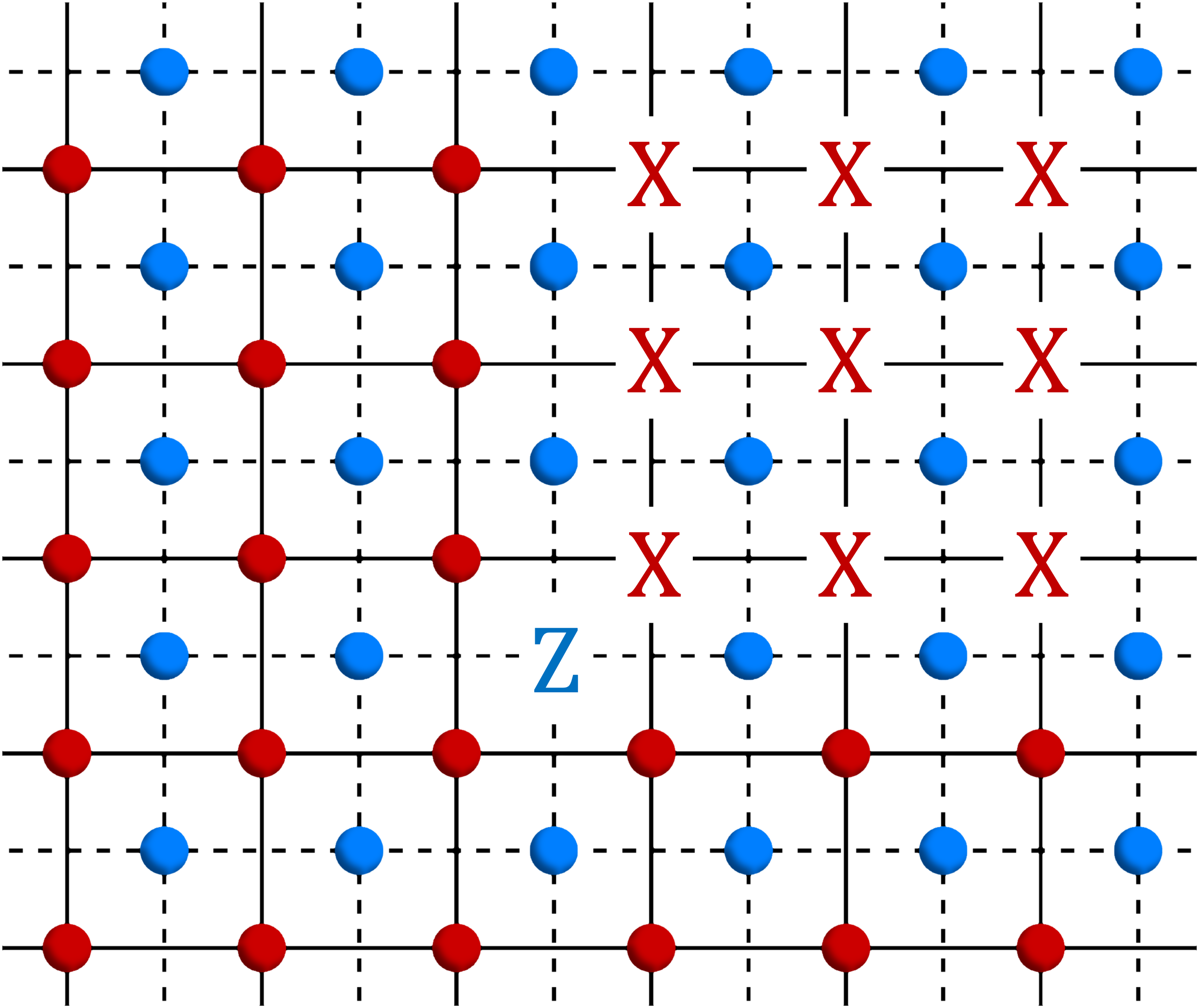}
    \caption{}
    \end{subfigure}
\caption{(a) Each site resides a red qubit, and each face resides a blue qubit. (b, c) $\text{KT}^{(2)}$ transformation acting on a single $Z$ operator is given by a product of Pauli $X$ operators in a light cone.}
\label{fig:three graphs}
\end{figure*}

In two dimensions, SSPT phases with symmetry group $G$ are classified by~\cite{devakul2018classification}
\begin{align}
    \mathcal{C}[G]\equiv H^2(G^2,U(1))/\left(H^2(G,U(1))\right)^3.
\end{align}
\red{As a remark by~\cite{devakul2018classification}, this is the classification of strong SSPT phases with symmetry group $G$. There is also a notion of weak SSPT phases which is composed of decoupled 1D SPTs. We will not consider weak SSPT phases in our analysis}. In our case $G=\mathbb{Z}_2\times\mathbb{Z}_2$ and $\mathcal{C}[\mathbb{Z}_2\times\mathbb{Z}_2]=\mathbb{Z}_2\times\mathbb{Z}_2\times \mathbb{Z}_2$. There are three generators for the SSPT. To enumerate them, 
let us consider a two-dimensional lattice consisting of two square sublattices. We color the sublattices with red and blue as shown in Fig.~\ref{fig:2drblattice}. We denote the vertices and plaquettes in each red and blue lattice by $v^r$ and $p^r$, and by $v^b$ and $p^b$, respectively. Note that the vertex of the red sublattice is effectively the plaquette of the blue sublattice and vice versa. The three generators for the SSPT are 1) $\mathbb{Z}_2\times\mathbb{Z}_2$ SSPT with cluster entangler between adjacent red and blue sites, 2) $\mathbb{Z}_2$ SSPT (given in section~\ref{sec:z2subsymmetricmodel}) on the red lattice, and  3) $\mathbb{Z}_2$ SSPT on the blue lattice.

Here, we will describe the first generator: $\mathbb{Z}_2\times\mathbb{Z}_2$ SSPT with cluster entangler between adjacent red and blue sites and the KT transformation that maps between this phase and two copies of the SSSB phase. The last two generators and their KT transformation are given in subsection~\ref{subsec:KTz_2SSPTtoSSSB}. For a general element in the classification $\mathbb{Z}_2\times\mathbb{Z}_2\times\mathbb{Z}_2$, in principle, we could construct a similar KT transformation that maps between this particular phase to copy/copies of SSSB phases. 

Let us consider the two-dimensional cluster state, with the red and blue sites entangled, which is the ground state of the first two terms in the following Hamiltonian
\begin{widetext}
\begin{align}
\begin{split}
    \bm H_{\text{2dcluster}}&=-\sum_{v_r}X_{v^r}\prod_{v^b\in \partial p^b}Z_{v^b}-\sum_{v^b}X_{v^b}\prod_{v^r\in \partial p^r}Z_{v^r} -\lambda \sum_{v^r} X_{v^r}-\lambda\sum_{v^b} X_{v^b},
\end{split}
\label{eq:2dcluster}
\end{align}
\end{widetext}
where we used the identification $p^b=v^r$ and $p^r=v^b$, and the last two terms are external fields to tune the system away from the cluster-state point. 
We assume periodic boundary conditions along both the $x$-axis and $y$-axis of the two-dimensional lattice. Note that all our results also carry over to the case of an infinite lattice. We define the two-dimensional KT transformation as
\begin{align} \label{eq:KT-2d-Z2Z2}
    \text{KT}^{(2)}\equiv\mathbf{D}_b^{(2)\dagger}\mathbf{D}_r^{(2)\dagger} T^{(2)}\bm \mathbf{D}_r^{(2)}\mathbf{D}_b^{(2)} ,
\end{align}
where $\textbf{D}_r^{(2)}$ and $\textbf{D}_b^{(2)}$ denote the KW transformation in two dimensions as defined in Eq.~\eqref{eq:D^2} for the red and blue sublattices. 
$T^{(2)}$ denotes the cluster entangler between red and blue sublattices and can be written explicitly as $T^{(2)}=\prod_{p}\prod_{v\in \partial p}CZ_{v,p}$.
Explicitly, $\text{KT}^{(2)}$ transformation is given by
\begin{subequations}
\begin{align}
    &\text{KT}^{(2)}X^r_{i,j}=X^r_{i,j}\text{KT}^{(2)}\,\\
    \begin{split}
        &\text{KT}^{(2)} X^b_{i+\frac{1}{2},j+\frac{1}{2}}Z^r_{i,j}Z^r_{i+1,j}Z^r_{i,j+1}Z^r_{i+1,j+1}\\
    &\qquad=Z^r_{i,j}Z^r_{i+1,j}Z^r_{i,j+1}Z^r_{i+1,j+1}\text{KT}^{(2)}
    \end{split}\,
\end{align}
\end{subequations}
and similar equations for $r\leftrightarrow b$. 
\red{Readers may wonder definitions of the order parameter of the SSSB phase and the string order parameter, and how they are related by the KT transformation. We discuss this in Appendix~\ref{ref:order-param}.}
The action of $\text{KT}^{(2)}$ on a single $Z$ operator on the red sublattice is 
\begin{align}
    \begin{split}
        \text{KT}^{(2)}Z^r_{i,j}&=Z^r_{i,j}\prod_{k=1}^{i}\prod_{l=1}^{j}X^b_{k-\frac{1}{2},l-\frac{1}{2}}\text{KT}^{(2)'},
    \end{split}
    \label{eq:KT2singleZr}
\end{align}
where 
\begin{align}
    \text{KT}^{(2)'}\equiv\mathbf{D}_b^{(2)\dagger}\mathbf{D}_r^{(2)'\dagger}T^{(2)}\bm \mathbf{D}_r^{(2)'}\mathbf{D}_b^{(2)} ,
\end{align}
with 
\begin{align}
    \textbf{D}_r^{(2)'}&\equiv\mathbf{\tilde{P}}^{\red{(2)}'}_{\red{r}}\tilde{\mathbf{D}}_{x,r}\red{\mathbf{H}^{\otimes(2)}_{r}}\tilde{\mathbf{D}}_{y,r}\mathbf{\tilde{P}}^{\red{(2)}}_{\red{r}},
    \label{eq:Dr^2}
\end{align}
and
\begin{subequations}
\begin{align}
    \mathbf{\tilde{P}}^{\red{(2)}}_{\red{r}}&\equiv\prod_{l\neq j}\frac{(1+\eta^x_l)}{2}\prod_{k\neq i}\frac{(1+\eta^y_k)}{2}\frac{(1-\eta^x_j)}{2}\frac{(1-\eta^y_i)}{2}.
\end{align}
\begin{align}
    \begin{split}
        \mathbf{\tilde{P}}^{\red{(2)}'}_{\red{r}}&\equiv\prod_{l=1}^j\frac{\left(1-\eta^x_l\right)}{2}\prod_{l=j+1}^{L_y}\frac{\left(1+\eta^x_l\right)}{2}\prod_{k\neq 1,i,i+1}\frac{\left(1+\eta^y_k\right)}{2}\\
     &\quad\times\frac{(1-\eta^y_i)}{2}\frac{(1-\eta^y_{i+1})}{2}\frac{(1+(-1)^j\eta^y_1)}{2},
    \end{split}
    \end{align}
\end{subequations}
$\tilde{\mathbf{D}}_{x,r}$ and $\tilde{\mathbf{D}}_{y,r}$ in Eq.~\eqref{eq:Dr^2} denote the operator $\tilde{\mathbf{D}}_{x}$ \red{and $\tilde{\mathbf{D}}_{y}$ on the red sublattice and $\mathbf{H}^{\otimes(2)}_r$ denote the tensor product of Hadamard operator on all sites of the red sublattice}.  $\text{KT}^{(2)}$ acting on a single $Z$ operator on the blue sublattice is 
\begin{align}
    \begin{split}
        \text{KT}^{(2)}Z^b_{i+\frac{1}{2},j+\frac{1}{2}}&=Z^b_{i+\frac{1}{2},j+\frac{1}{2}}\prod_{k=i+1}^{L_x-1}\prod_{l=j+1}^{L_y-1}X^r_{k,l}\text{KT}^{(2)''},
    \end{split}
    \label{eq:KT2singleZb}
\end{align}
where 
\begin{align}
    \text{KT}^{(2)''}\equiv\mathbf{D}_b^{(2)'\dagger}\mathbf{D}_r^{(2)\dagger}T^{(2)}\bm \mathbf{D}_r^{(2)}\mathbf{D}_b^{(2)'} ,
\end{align}
with 
\begin{align}
    \textbf{D}_b^{(2)'}&\equiv\mathbf{\tilde{P}}^{\red{(2)}'}_{\red{b}}\tilde{\mathbf{D}}_{x,b}\red{\mathbf{H}^{\otimes(2)}_{b}}\tilde{\mathbf{D}}_{y,b}\mathbf{\tilde{P}}^{\red{(2)}}_{\red{b}},
    \label{eq:Db^2}
\end{align}
and   
\begin{widetext}
\begin{subequations}
\begin{align}
    \begin{split}
        \mathbf{\tilde{P}}^{\red{(2)}'}_{\red{b}}&\equiv\prod_{l=1}^{j+1\red{-1}}\frac{\left(1-\eta^x_{l+\frac{1}{2}}\right)}{2}\prod_{l=j+1}^{L_y}\frac{\left(1+\eta^x_{l+\frac{1}{2}}\right)}{2}\prod_{k\neq 1,i,i+1}\frac{\left(1+\eta^y_{k+\frac{1}{2}}\right)}{2}\frac{(1-\eta^y_{i+\frac{1}{2}})}{2}\frac{(1-\eta^y_{i+\frac{3}{2}})}{2}\frac{(1+(-1)^{j+1}\eta^y_{\frac{1}{2}})}{2},
    \end{split}\\
    \begin{split}
        \mathbf{\tilde{P}}^{\red{(2)}}_{\red{b}}&\equiv\prod_{l\neq j}\frac{(1+\eta^x_{l+\frac{1}{2}})}{2}\prod_{k\neq i}\frac{(1+\eta^y_{k+\frac{1}{2}})}{2}\frac{(1-\eta^x_{j+\frac{1}{2}})}{2}\frac{(1-\eta^y_{i+\frac{1}{2}})}{2}.
    \end{split}
\end{align}
\end{subequations}
\end{widetext}
$\tilde{\mathbf{D}}_{x,b}$ and $\tilde{\mathbf{D}}_{y,b}$ in Eq.~\eqref{eq:Db^2} denote the operator $\tilde{\mathbf{D}}_{x}$ and $\tilde{\mathbf{D}}_{y}$ on the blue sublattice \red{and $\red{\mathbf{H}^{\otimes(2)}_b}$ denote the tensor product of Hadamard operator on all sites of the blue sublattice}. The subsystem symmetry generators on the blue sublattice are $\eta_{j+\frac{1}{2}}^x=\prod_{i=0}^{L_x-1}X_{i+\frac{1}{2},j+\frac{1}{2}}$ and $\eta_{i+\frac{1}{2}}^y=\prod_{j=0}^{L_y-1}X_{i+\frac{1}{2},j+\frac{1}{2}}$. The structure of the membrane operator in Eq.~\eqref{eq:KT2singleZr} and Eq.~\eqref{eq:KT2singleZb}
(see Fig.~\ref{fig:three graphs}b \& c) for the derived $\text{KT}^{(2)}$ transformation resembles the structure given in Ref.~\cite{doherty2009identifying}, except for the rotation of the lattice by $45^\circ$. 
Their mapping of a Pauli $Z$ operator involves a product of Pauli operators supported on a quadrant (light cone) relative to the original Pauli $Z$ operator, where the original ones on the red and blue lattices get mapped to the opposite quadrants.
Similar to theirs, we find a light cone structure in our transformation. We also find that the composition of two $\text{KT}^{(2)}$ transformations gives
\begin{align}
    (\text{KT}^{(2)})^2\propto\,\mathbf{P}^{\red{(2)}}_r\mathbf{P}^{\red{(2)}}_b,
\end{align}
 $\mathbf{P}^{\red{(2)}}_r$ and $\mathbf{P}^{\red{(2)}}_b$ are the respective projections onto the \red{subspaces of the} Hilbert spaces of the red and blue sublattices, which are separately subsystem symmetric.

Similar to the remark in the 1D case, there are different ways to implement a valid KT transformation. We refer the readers to appendix~\ref{sec:TST} for explicit calculations of another choice.

\subsection{Higher dimensions}
Here, we will generalize all the discussions we had for two dimensions to higher dimensions. We consider $d$ dimensional lattice consisting of two hypercubic sublattices. These sublattices are dual to each other. Again, we color them red and blue. Let us denote the vertices and hypercubes of the red and blue sublattices by $v^r$,$c^r$ and $v^b$,$c^b$ respectively. A vertex of the red sublattice is effectively a hypercube of the blue sublattice. 
Let us consider the $d$-dimensional entangler on the red and blue hypercubic sublattices for the cluster state described by the Hamiltonian
\begin{widetext}
\begin{align}
\begin{split}
    \bm H_{\text{d-dimcluster}}&=-\sum_{v_r}X_{v^r}\prod_{v^b\in \partial c^b}Z_{v^b}-\sum_{v^b}X_{v^b}\prod_{v^r\in \partial c^r}Z_{v^r} -\lambda\sum_{v^r} X_{v^r}-\lambda\sum_{v^b} X_{v^b},
\end{split}
\label{eq:dDcluster}
\end{align}
\end{widetext}
where we used the identification $c^b=v^r$ and $c^r=v^b$; 
see Fig.~\ref{fig:Z2Z2_hypercluster} for an illustration of Hamiltonian terms. 
We assume periodic boundary conditions along all the $d$ axes. We define the $d$ dimensional KT transformation
\begin{align}
    \text{KT}^{(d)}\equiv\mathbf{D}_b^{(d)\dagger}\mathbf{D}_r^{(d)\dagger}T^{(d)}\mathbf{D}_r^{(d)}\mathbf{D}_b^{(d)},
\end{align}
where $\mathbf{D}_r^{(d)}$ and $\mathbf{D}_b^{(d)}$ denote the KW transformation in $d$ dimensions defined in Eq.~\eqref{eq:D^d} for red and blue sublattices. $T^{(d)}$ is the cluster entangler between the red and blue sublattices. Explicitly, $\text{KT}^{(d)}$ transformation is given by
\begin{subequations}
\begin{align}
    \text{KT}^{(d)}X_{v^r}&=X_{v^r}\text{KT}^{(d)},\\
    \text{KT}^{(d)}X_{v^b}\prod_{v^r\in \partial c^r}Z_{v^r}&=\prod_{v^r\in \partial c^r}Z_{v^r}\text{KT}^{(d)},
\end{align}
\end{subequations}
 By $r\leftrightarrow b$, we get another similar set of equations. The action of $\text{KT}^{(d)}$ on a single $Z$ operator can be worked out similarly in the two-dimensional case, and we find that the light cone structure is in opposite directions on the two sublattices~\cite{you2018subsystem}.  
The composition of two $\text{KT}^{(d)}$ transformations is \red{proportional to a projection}
\begin{align}
    (\text{KT}^{(d)})^{2}\red{\propto}\,\mathbf{P}_r^{(d)}\mathbf{P}_b^{(d)} .
\end{align}
 $\mathbf{P}_r^{(d)}$ and $\mathbf{P}_b^{(d)}$ are projections onto subsystem symmetric subspace of the Hilbert spaces of red and blue sublattices. In appendix~\ref{app:Measurement-based}, we describe how to implement the KT transformation in a short-depth operation.\\\\

\section{Composition of operators beyond one dimension}\label{sec:Comptwoandhigher}
\subsection{Two dimensions}
  We consider the composition of Kennedy-Tasaki and Kramers-Wannier transformation in two dimensions shown with the lattice given in Fig.~\ref{fig:2drblattice}. This is a similar analysis as we did in subsection~\ref{subsec:Compoperators}.
  \begin{align}
      \mathbf{D}_{SSPT}^{(2)}\equiv\text{KT}^{(2)\dagger}\:\mathbf{D}_{\red{r}}^{(2)}\red{\mathbf{D}_{\red{b}}^{(2)}}\:\text{KT}^{(2)}.
  \end{align}
  It satisfies the following relations
  \begin{widetext}
  \begin{subequations}
      \begin{align}
      \mathbf{D}_{SSPT}^{(2)}X^r_{i,j}=X^b_{i+\frac{1}{2},j+\frac{1}{2}}Z^r_{i,j}Z^r_{i+1,j}Z^r_{i,j+1}Z^r_{i+1,j+1}\mathbf{D}_{SSPT}^{(2)},\\
      \mathbf{D}_{SSPT}^{(2)}X^b_{i+\frac{1}{2},j+\frac{1}{2}}Z^r_{i,j}Z^r_{i+1,j}Z^r_{i,j+1}Z^r_{i+1,j+1}=X^r_{i+1,j+1}\mathbf{D}_{SSPT}^{(2)}.
  \end{align}
  \label{eq:DSSPT2transformation}
   \end{subequations}
\end{widetext}
$\mathbf{D}_{SSPT}^{(2)}$ is a symmetry of the 2D cluster state Hamiltonian Eq.~\eqref{eq:2dcluster} with $\mathbb{Z}_2\times\mathbb{Z}_2$ subsystem symmetry. Moreover, it squares to a translation times a projection as in the case of Kramers-Wannier transformation Eq.~\eqref{eq:TP2d},
  \begin{align}
      \left(\mathbf{D}_{SSPT}^{(2)}\right)^2\propto\,\mathbf{T}_{(1,1)\red{,r}}\red{\mathbf{T}_{(1,1),b}}\mathbf{P}^{\red{(2)}}_{\red{r}}\red{\mathbf{P}^{\red{(2)}}_{\red{b}}}.
  \end{align}
\red{The translations $\mathbf{T}_{(1,1),r}$ and $\mathbf{T}_{(1,1),b}$  act on both the red and blue sublattice by a diagonal unit shit on the respective sublattices.}
\red{We note that the transformations Eq.~\eqref{eq:DSSPT2transformation} are equivalently achieved by the 
\blue{controlled-Z}
operator $\mathcal{U}_{CZ}^{(2)}=\prod_{v^r}\prod_{v^b=p^r\in\partial^*v_r}CZ_{v^r,p^r}$\blue{ with $\partial^* v_r$ a set of plaquettes that contain $v_r$}. $\mathcal{U}_{CZ}^{(2)}$ is a unitary and squares to 1. }

\subsection{Higher dimensions}
Similarly, we can consider the composition of Kennedy-Tasaki and Kramers-Wannier in higher dimensions.
  \begin{align}
      \mathbf{D}_{SSPT}^{(d)}\equiv\text{KT}^{(d)\dagger}\:\mathbf{D}_{\red{r}}^{(d)}\red{\mathbf{D}_{\red{b}}^{(d)}}\:\text{KT}^{(d)}.
  \end{align}
  It satisfies the following relations
  \begin{subequations}
      \begin{align}
     \mathbf{D}_{SSPT}^{(d)}X_{v^r}=X_{v^b}\prod_{v^r\in \partial c^r}Z_{v^r}\mathbf{D}_{SSPT}^{(d)}\label{eq:Xvr},\\
     \mathbf{D}_{SSPT}^{(d)}X_{v^b}\prod_{v^r\in \partial c^r}Z_{v^r}=X_{v^r+d}\mathbf{D}_{SSPT}^{(d)},
  \end{align}
  \label{eq:DSSPTdtransformation}
\end{subequations}
  where $d$ denotes the direction $(1,1,...,1)$ and is the diagonal vector. The vertices $v^r$ and $v^b$ \red{appearing as subscripts of Pauli $X$ operators} in Eq.~\eqref{eq:DSSPTdtransformation} are related by a half diagonal translation $(\frac{1}{2},\frac{1}{2},...,\frac{1}{2})$, i.e., $v^b=v^r+(\frac{1}{2},\frac{1}{2},...,\frac{1}{2})$.
  Hence, $\mathbf{D}_{SSPT}^{(d)}$ is a symmetry of the d-D cluster state Hamiltonian Eq.~\eqref{eq:dDcluster} with $\mathbb{Z}_2\times \mathbb{Z}_2$ subsystem symmetry. Again, it squares to diagonal translation times a projection \red{up to a normalization factor}
  \begin{align}
     (\mathbf{D}_{SSPT}^{(d)})^2\red{\propto}\,\mathbf{T}_{(1,...,1),\red{r}}\red{\mathbf{T}_{(1,...,1),b}}\mathbf{P}^{\red{(d)}}_{\red{r}}\red{\mathbf{P}^{\red{(d)}}_{\red{b}}}.
  \end{align}
  \red{The translations $\mathbf{T}_{(1,...,1),r}$ and $\mathbf{T}_{(1,...,1),b}$  act on both the red and blue sublattice by a diagonal unit shit on the respective sublattices.}
 \red{We note that the transformations Eq.~\eqref{eq:DSSPTdtransformation} are equivalently achieved by the  
 \blue{controlled-Z}
 operator $\mathcal{U}_{CZ}^{(d)}=\prod_{v^r}\prod_{v^b=c^r\in\partial^*v_r}CZ_{v^r,c^r}$ 
 \blue{ with $\partial^*v_r$ a set of hypercubes that contain $v_r$}. $\mathcal{U}_{CZ}^{(d)}$ is a unitary and square to 1.}
\begin{widetext}
\begin{center}
\begin{figure}[h!]
    \includegraphics[scale=1]{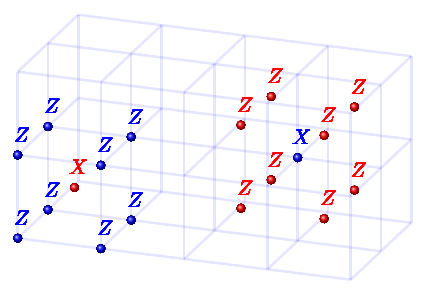}
    \caption{Hamiltonian terms in the $\mathbb{Z}_2 \times \mathbb{Z}_2$ cluster state on the three-dimensional lattice. Here, we draw the blue lattice as a background. The red lattice sites reside at the centers of the blue cubes. (left) $X_{v^r}\prod_{v^b\in \partial c^b}Z_{v^b}$, (right) $X_{v^b}\prod_{v^r\in \partial c^r}Z_{v^r}$.}
    \label{fig:Z2Z2_hypercluster}
\end{figure}
\end{center}
\end{widetext}

\section{Non-invertible symmetry in $\mathbb{Z}_2$ subsystem symmetric models}\label{sec:z2subsymmetricmodel}

In this section, we construct a non-invertible symmetry in a $\mathbb{Z}_2$ subsystem symmetric model in two dimensions, which exhibits an SSSB order in the limit of large coupling. 
Moreover, we construct a map from a $\mathbb{Z}_2$ SSPT cluster state to a trivial state, as well as that from the SSPT cluster state to the SSSB state.
(Note that, unlike in the previous sections, the symmetry in the SSPT is not $\mathbb{Z}_2 \times \mathbb{Z}_2$ but $\mathbb{Z}_2$ instead.)
We study the transformation in two dimensions in detail, but the construction generalizes to arbitrary spatial dimensions. 
At the end of the section, we briefly discuss the generalization to three dimensions and higher. 

\subsection{Kramers-Wannier for SSSB Hamiltonian}

Consider a Hamiltonian, which we may call the double-plaquette Ising model (DPIM),
\begin{widetext}
\begin{align}
H_\text{DPIM}(\lambda) 
&= - \sum_{i,j} 
\Big( Z_{i+1,j+1}
Z_{i+1,j}
Z_{i,j+1}
Z_{i-1,j}
Z_{i,j-1}
Z_{i-1,j-1} + \lambda X_{i,j}\Big) = - \sum_{i,j}  
\left[
\begin{array}{ccc}
 & Z & Z \\
Z&   & Z \\
Z& Z &
\end{array}
+ \lambda X  \right]
. 
\label{eq:sixZIsing}
\end{align}
\end{widetext}
For simplicity, we consider the model on a square lattice with $L_x =L_y=L$.
One can think of the first term as the product of two plaquette Ising terms shifted in both $x$ and $y$ directions by one. 
The model is symmetric under a set of unitary transformations that represents spin flips along rigid lines in $x$, $y$, and diagonal directions:
{\allowdisplaybreaks
\begin{subequations}
\begin{align}
\eta^x_j &\equiv \prod_{i = 1 }^L X_{i,j} \quad (j \in \{1, ...,L\}), \\
\eta^y_i &\equiv \prod_{j = 1 }^L X_{i,j} \quad (i \in \{1, ...,L\}), \\
\eta^\text{diag}_k &\equiv \prod_{\ell = 1 }^L X_{ \ell,[\ell+k]_L} \quad (k \in \{1, ...,L\}), 
\end{align}
\label{eq:2dsubsystemwithdiag}
\end{subequations} }
where $[\bullet]_L$ denotes the entry mod $L$.
It is a subsystem $\mathbb{Z}_2$ symmetry with a constraint $\prod_{j=1}^L\eta^x_j = \prod_{i=1}^L\eta^y_i = \prod_{k=1}^L\eta^\text{diag}_k$. 
In what follows, we consider gauging all of the above symmetries. 
To do so, we employ the sequential circuit approach and gauge each $\mathbb{Z}_2$ line symmetry one by one, which implements the 1d Kramers-Wannier transformation on it. 
It is natural to expect that the whole map executes a self-duality, and indeed, the Hamiltonian $H_\text{DPIM}$ is mapped to itself with $\lambda$ moved to the $Z$ term.

Now, we define the sequential circuit.
We use the operators $\tilde{\mathbf{D}}_x$ and $\tilde{\mathbf{D}}_y$, already defined in Eq.~\eqref{Dx-def} and Eq.~\eqref{Dy-def}, respectively. 
We introduce another unitary,
\begin{align}
\tilde{\mathbf{D}}_\text{diag}\equiv\prod_{k=1}^{L}d_k \label{Ddiag-def} ,
\end{align}
with
\begin{align}
d_k\equiv&\left(\prod_{\ell=1}^{L-1}e^{i\frac{\pi}{4}X_{\ell,[\ell+k]_L}}e^{i\frac{\pi}{4}Z_{\ell,[\ell+k]_L}Z_{\ell+1,[\ell+k+1]_L}}\right) \nonumber \\
&\quad \times e^{i\frac{\pi}{4}X_{L,[L+k]_L}} ,
\end{align}
where the ordering is understood as before; as $\ell$ increases, we go to the right in the product.
We use a projector to absorb unwanted $\eta$'s that get attached to boundary terms in the Hamiltonian upon transformations by unitaries:
\begin{align}
\mathbf{P}_\text{DPIM}
\equiv \prod_{j = 1}^L \frac{1+\eta^x_j}{2}
\prod_{i = 1}^L \frac{1+\eta^y_i}{2}
\prod_{k = 1}^L \frac{1+\eta^\text{diag}_k}{2}  .
\end{align}
Then we define the duality operator as
\begin{align}
\mathbf{D}_\text{DPIM}
\equiv 
\mathbf{P}_\text{DPIM}
\tilde{\mathbf{D}}_y
\red{\mathbf{H}^{\otimes(2)}}
\tilde{\mathbf{D}}_x
\red{\mathbf{H}^{\otimes(2)}}
\tilde{\mathbf{D}}_\text{diag}
\mathbf{P}_\text{DPIM} ,
\label{eq:DDPIM}
\end{align}
where $\red{\mathbf{H}^{\otimes(2)}}$ is the simultaneous Hadamard transformation on all the qubits, as before.
In the bulk, the transformation occurs as below:
\begin{widetext}
\begin{align}
\begin{array}{ccc}
 &  &  \\
 &   &  \\
\boxed{X}&  &
\end{array}
\overset{\tilde{\mathbf{D}}_\text{diag}}{\longrightarrow}
\begin{array}{ccc}
 &  &  \\
 & Z  &  \\
\boxed{Z}&  &
\end{array}
\overset{\red{\mathbf{H}^{\otimes(2)}}}{\longrightarrow}
\begin{array}{ccc}
 &  &  \\
 & X  &  \\
\boxed{X}&  &
\end{array}
\overset{\tilde{\mathbf{D}}_{x}}{\longrightarrow}
\begin{array}{ccc}
 &  &  \\
 & Z  & Z \\
\boxed{Z}& Z &
\end{array}
\overset{\red{\mathbf{H}^{\otimes(2)}}}{\longrightarrow}
\begin{array}{ccc}
 &  &  \\
 & X  & X \\
\boxed{X}& X &
\end{array} 
\overset{\tilde{\mathbf{D}}_{y}}{\longrightarrow}
\begin{array}{ccc}
 & Z & Z \\
Z&   & Z \\
\boxed{Z}& Z &
\end{array}  \ , 
\end{align}
\end{widetext}
where we have added boxes to indicate the same location on the lattice.
The transformations form the following algebra:
\begin{subequations}
\begin{align}
&[H_\text{DPIM}(\lambda), \eta] = 0  , \\
& H_\text{DPIM}(\lambda=1) \mathbf{D}_\text{DPIM} =  \mathbf{D}_\text{DPIM} H_\text{DPIM}(\lambda=1) ,  \\
&  \eta \mathbf{D}_\text{DPIM}  = \mathbf{D}_\text{DPIM}  \eta = \mathbf{D}_\text{DPIM} , 
\end{align}
\end{subequations}
for $\eta \in \{\eta^x_j, \eta^y_i, \eta^\text{diag}_k\}$; in particular at the self-duality point $\lambda =1$, we get an algebra involving a non-invertible symmetry.
Moreover, \red{$\mathbf{D}_\text{DPIM}$ squares to a diagonal translation by two units and a projection up to an overall normalization}
\begin{align}
\left(\mathbf{D}_\text{DPIM} \right)^2 \red{\propto}\, \mathbf{T}_{(2,2)} \mathbf{P}_\text{DPIM} , 
\end{align}
where $\mathbf{T}_{(2,2)}$ is the translation by $(2,2)$.
Note, however, that one could set the opposite ordering for the diagonal part of the unitary.
Then such a KW operator $\mathbf{D}'_\text{DPIM}$ defined with it would obey $\left(\mathbf{D}'_\text{DPIM} \right)^2 \red{\propto}\, \mathbf{P}_\text{DPIM}$, so the translation is not an essential feature in this case.
We also comment that we can implement the KW transformation in a short-depth operation (including measurement) in appendix~\ref{app:Measurement-based}.

\subsection{Kennedy-Tasaki for SSSB and SSPT}\label{subsec:KTz_2SSPTtoSSSB}

In the literature~\cite{you2018subsystem,devakul2018classification}, the following Hamiltonian at $\lambda =0$ is known to host a $\mathbb{Z}_2$ SSPT ground state:
\begin{widetext}
\begin{align}
H_\text{SSPT}(\lambda) 
&= - \sum_{i,j} 
\Big( X_{i,j}Z_{i+1,j+1}
Z_{i+1,j}
Z_{i,j+1}
Z_{i-1,j}
Z_{i,j-1}
Z_{i-1,j-1} + \lambda X_{i,j}\Big) = - \sum_{i,j}  
\left[
\begin{array}{ccc}
 & Z & Z \\
Z& X & Z \\
Z& Z &
\end{array}
+ \lambda X  \right]
. 
\label{eq:sixZSSPT}
\end{align}
\end{widetext}
This Hamiltonian can be obtained from a symmetry breaking of the 2d cluster model in section~\ref{sec:2dcluster}, which is the 2d $\mathbb{Z}_2\times \mathbb{Z}_2$ SSPT for horizontal and vertical line-like symmetry on square lattice~\cite{devakul2018classification}. Namely, if we insert a term 
\beq
    -g\sum \begin{matrix}
        & Z_r\\
        Z_b& 
    \end{matrix},
\eeq
into the Hamiltonian in Eq.~\eqref{eq:2dcluster}, and tune $g\rightarrow\infty$, the symmetry will then be broken into the diagonal subgroup $\mathbb{Z}_2\times \mathbb{Z}_2\rightarrow\mathbb{Z}_2^{diag}$, as shown in apppendix~\ref{sec:symmetrybreaking}. 

There is an obvious cluster-state entangler which maps from $X_{i,j}$ to the first term, which we denote by $T_\text{SSPT}$,
\begin{align}
\begin{array}{ccc}
 &  &  \\
 & \boxed{X}  &  \\
&  &
\end{array} 
\overset{T_\text{SSPT}}{\longleftrightarrow}
\begin{array}{ccc}
 & Z & Z \\
Z&  \boxed{X} & Z \\
Z& Z &
\end{array}.  
\end{align}
We remark that in the literature, the cluster state described by the above stabilizer is seen as an SSPT state protected by $\{\eta^x_j, \eta^y_i\}$, see e.g., Refs.~\cite{you2018subsystem,devakul2018classification}. 
In our current context, on the other hand, it would also be natural to view this as an SSPT order protected by $\{\eta^x_j, \eta^y_i, \eta^\text{diag}_k\}$. 

Indeed, we have the following mapping:
\begin{subequations}
\begin{align}
\begin{array}{cccc}
 & & & \\ 
 & Z &Z & \\
 Z& \boxed{X}  &Z &  \\
Z&  Z& &
\end{array} 
\overset{\mathbf{D}_\text{DPIM}}{\longleftrightarrow}
\begin{array}{cccc}
&  & Z & Z \\
& Z&  X & Z \\
& \boxed{Z}& Z & \\
& &  & 
\end{array}  ,  
\\
\begin{array}{ccc}
&  &  \\
&  &  \\
\boxed{X}& \white{X} & \white{X}  
\end{array} 
\overset{\mathbf{D}_\text{DPIM}}{\longleftrightarrow}
\begin{array}{ccc}
 & Z & Z \\
Z&   & Z \\
\boxed{Z}& Z &
\end{array}.
\end{align}
\end{subequations}
Now note that the map $\mathbf{D}_\text{DPIM}$ preserves locality for symmetric operators (those composed of single $X$ and the product of six $Z$'s), and it also preserves a gap. 
The property $\eta \mathbf{D}_\text{DPIM} = \mathbf{D}_\text{DPIM}$ means that this is a gauging map~\cite{levin2012braiding, yoshida2016topological,yoshida2017gapped, kubica2018ungauging}.
We expect that there is no finite-depth local unitary circuit connecting the symmetric ground states described by the two stabilizers on the right-hand side: the one being the cluster state (short-range entangled) and the other being SSSB with long-range order. 
Then one cannot have a symmetric (under $\{\eta^x_j, \eta^y_i, \eta^\text{diag}_k\}$) finite-depth local unitary connecting the states described by the two stabilizers on the left-hand side. 
Since $X$ is the stabilizer for the trivial symmetric state, the other one has to belong to a non-trivial SSPT. 
Finally, we can define a Kennedy-Tasaki transformation:
\begin{align}
\text{KT}_{\mathbb{Z}_2}
\equiv \mathbf{D}_\text{DPIM}^{\dagger} T_\text{SSPT} \mathbf{D}_\text{DPIM} , 
\end{align}
which implements the transformation between the SSPT and SSSB phases,
\begin{align}
\begin{array}{ccc}
 & Z &Z  \\
 Z& X  &Z  \\
\boxed{Z}&  Z&
\end{array} 
\overset{\text{KT}_{\mathbb{Z}_2}}{\longleftrightarrow}
\begin{array}{ccc}
 & Z & Z \\
Z&   & Z \\
\boxed{Z}& Z &
\end{array} .
\label{eq:KTsinglez2}
\end{align}
Equivalently we can define
\begin{align}
\text{KT}_{\mathbb{Z}_2}
\equiv T_\text{SSPT} \mathbf{D}_\text{DPIM} T_\text{SSPT}  , 
\end{align}
and implements the same transformation in Eq.~\eqref{eq:KTsinglez2} with a diagonal unit shift.
\subsection{Three dimensions and higher}

\begin{figure}
   \includegraphics[scale=1]{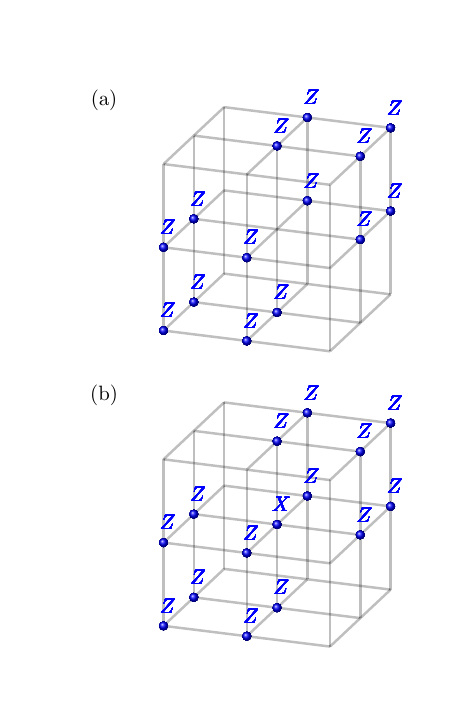}
   \caption{(a) A Hamiltonian term in the double hypercube Ising model in three dimensions. Note that there is no $Z$ operator on the vertex at the center. (b) A Hamiltonian term in the $\mathbb{Z}_2$ SSPT cluster state in three dimensions.}
   \label{fig:DCIM}
\end{figure}

We conclude this section by giving a general picture in higher dimensions. 
The model we consider as a generalization of DPIM is a model that we shall call a double hypercube Ising model, whose multi-body term is simply the product of shifted hypercube terms $\prod_{v\in\partial c}Z_{v}$ in Eq.~\eqref{TFHCI}; 
see Fig.~\ref{fig:DCIM}(a) for an illustration of three dimensions.
Inherited from the parent Hamiltonian, the model is symmetric under spin flips along rigid lines in every coordinate direction. 
Due to the shifted product, the model is also symmetric under the spin flips along any rigid diagonal line pointing in the direction $(1,1,\dots,1)$. 
Then, the same story in two dimensions can be generalized to three and higher dimensions. 

In three dimensions, for example, we have a double cube Ising model (DCIM), whose Ising term is a product of $Z$ operators at fourteen points, {\it i.e.,} the product of eight $Z$'s at the corners of two cubes with one overlap. 
The sequential circuit is 
\begin{align}
\begin{split}
\mathbf{D}_\text{DCIM} 
&\equiv \mathbf{P}_\text{DCIM}
\tilde{\mathbf{D}}_z
\red{\mathbf{H}^{\otimes(3)}}
\tilde{\mathbf{D}}_y
\red{\mathbf{H}^{\otimes(3)}}\times\\
&\qquad\qquad\tilde{\mathbf{D}}_x
\red{\mathbf{H}^{\otimes(3)}}
\tilde{\mathbf{D}}_\text{diag}
\mathbf{P}_\text{DCIM}, 
\end{split}
\end{align}
with an appropriate definition of $\tilde{\mathbf{D}}_\text{diag}$\red{, $\red{\mathbf{H}^{\otimes(3)}}$} and $\mathbf{P}_\text{DCIM}$, where the latter imposes all the symmetry generators $\{\eta^x_j, \eta^y_i, \eta^z_m ,\eta^\text{diag}_k\}$ to be evaluated to unity for the ground states. 

On the other side of the duality web, there is also a cluster state whose stabilizer is given by one $X$ (sitting at the overlapping site of two diagonally neighboring cubes) and fourteen $Z$'s (sitting at the remaining corners),  which is produced by a combination of $CZ$ gates; see Fig.~\ref{fig:DCIM}(b). This 3D cluster state is a nontrivial 3D $\mathbb{Z}_2$ SSPT; see Ref.~\cite{tantivasadakarn2020searching} for other generators of 3D $\mathbb{Z}_2$ SSPT phases. 
The cluster-state entangler and the Kramers-Wannier duality operator $\mathbf{D}_\text{DCIM}$ form a web of dualities.

\section{Conclusion}\label{sec:conc}
In this paper, we have presented a higher-dimensional generalization of the non-invertible Kramers-Wannier duality symmetry on a lattice. The generalized hypercubic Ising models 
 with a transverse field exhibit non-invertible symmetry at the self-dual point. 
In addition to that, we have also presented a generalization of the Kennedy-Tasaki transformation in higher dimensions. 
In our examples involving $\mathbb{Z}_2\times\mathbb{Z}_2$ SSPT phases, under the KT transformation, the higher-dimensional cluster-state model with an external field decomposes into two copies of hypercubic transverse-field Ising models.
 The KT transformation derived in the main text is obtained by sandwiching the cluster-state entangler between the KW duality operator (on two sublattices) and its Hermitian conjugate. We have also derived an alternative KT transformation in the appendix~\ref{sec:TST}. Both of these variants of the KT transformation achieve the effect of taking a $\mathbb{Z}_2\times\mathbb{Z}_2$ model to two copies of SSSB models.
Our result generalizes the picture of the 1D KT transformation proposed by~\cite{li2023non} to higher dimensions. In addition to the $\mathbb{Z}_2\times\mathbb{Z}_2$ symmetry, we also discussed KW duality symmetry of DHCIM and used that to give a KT transformation that maps between the cluster model, which is $\mathbb{Z}_2$ SSPT, and one copy of DHCIM, which is in the SSSB phase.

While the KW operators in our construction require linear-depth circuits, they can also be implemented using a cluster-state entangler acting on the original and ancillary degrees of freedom, measurement on the original degrees of freedom, and then a feedforward correction, the whole combination of which is finite-depth (see appendix~\ref{app:Measurement-based}). It is, therefore, feasible to implement the KW and KT transformations on quantum devices, following the approaches in, e.g., Refs.~\cite{piroli2021quantum, tantivasadakarn2021long, lu2022measurement,iqbal2023topological}. On the other hand, in the case of 1D~\cite{seiberg2023majorana}, the linear-depth construction of the KW duality operator (on the same lattice) is a non-invertible symmetry arising from the anomalous translation symmetry in the Majorana fermion representation after gauging the global fermion parity. Hence, this construction clarifies the relationship between anomalies and non-invertible symmetry.  In appendix.~\ref{sec:fermionicdual}, we also discuss Majorana hypercube models and identify the exchange in the Majorana terms that gives rise to the KW duality in the corresponding Ising hypercube model.  The exchange action is not related to translation and, moreover, does not commute with the subsystem fermion parity. However, the physical meaning of this exchange symmetry is not yet clear.

The non-invertible symmetries would also have a generalization to a Kramers-Wannier transformation that gauges higher-form symmetries. 
In Ref.~\cite{roberts2017symmetry}, for example, gauging $\mathbb{Z}_2 \times \mathbb{Z}_2$ 1-form symmetry was considered using a mathematical map, which transforms the 3D cluster state (also called the Raussendorf-Bravyi-Harrington state~\cite{raussendorf2005long}) to itself and the product state to two copies of the 3D toric code; essentially, this is a higher-form symmetry generalization of the Kramers-Wannier transformation. 
We have confirmed that the parent Hamiltonians of the above-mentioned states map in the same corresponding way under the measurement-assisted construction.
It would be interesting to explore the KW transformation more broadly beyond this example. 
One could also define a higher-form generalization of the Kennedy-Tasaki transformation by composing the map with the 3D cluster state entangler, which in total brings an SPT state to some copies of SSB states with respect to the higher-form symmetry. 
It would be interesting to construct a circuit with a projector that realizes this transformation and study its algebra.

It is also natural to study the Kennedy-Tasaki transformations that map between the SSPT phase and the SSB phase in higher-dimensional models for more examples and other symmetry groups. 
Classification of SSPT phases beyond two dimensions is not extensively known yet.
Thus, it would be worthwhile to extend the classification of SSPT phases to higher dimensions and find a KT transformation that maps between all the SSPT to SSSB phases for symmetry groups beyond $\mathbb{Z}_2$ and $\mathbb{Z}_2\times\mathbb{Z}_2$.

\red{Recently, KT has been used to discuss SPT phases protected by non-invertible symmetry \cite{seifnashri2024cluster} in 1+1 dimensions. The distinct non-invertible SPT phases are characterized by various symmetry-breaking patterns of the dual symmetry that emerges after applying the KT transformation. It would be interesting to analyze such non-invertible SSPT phases in 2+1 dimensions. We speculate that the cluster states with $\mathbb{Z}_2\times \mathbb{Z}_2$ or $\mathbb{Z}_2$ subsystem symmetry would split into distinct SSPTs after imposing the non-invertible KW duality symmetry. However, we leave a comprehensive analysis for future exploration.}

The existence of non-invertible symmetry put constraints similar to the Lieb-Shultz-Mattis theorem on the low energy theory of one-dimensional lattice models: it is recently found in the case of 1D by Seiberg, Seifnashri, and Shao~\cite{seiberg2024non} that the system is either in a gapless phase or gapped phase with a three (or a multiple of three) degenerate ground states at the non-invertible symmetric point. 
Studying such LSM constraints with non-invertible symmetries beyond one dimension would be generally interesting.
\red{In two dimensions, the immediate question is whether lattice models with $\mathbb{Z}_2$ subsystem symmetry and the non-invertible symmetry $\mathbf{D}^{(2)}$ obey any such constraints.
Let us consider some possible $\mathbb{Z}_2$ subsystem symmetric phases on the two-dimensional torus.
\begin{itemize}
    \item According to the classification of strong SSPTs in two dimensions, there are
    \begin{itemize}
        \item the trivial SSPT ($\rm H_1$ in the first row),
        \item one non-trivial $\mathbb{Z}_2$ SSPT($\rm H_1$ in the second row).
    \end{itemize}   
    \item There is an SSSB  phase ($\rm H_2$ in the first row).
    \item There is a topological order, i.e., the Wen-plaquette model~\cite{wen2003quantum} ($\rm H_2$ in the second row).
\end{itemize}
The trivial SSPT is mapped to the SSSB phase under the Kramers-Wannier operator $\mathbf{D}^{(2)}$; see the first row of table \ref{tab:GSD2dim}. 
It is easy to see that the non-trivial $\mathbb{Z}_2$ SSPT that was discussed in section~\ref{subsec:KTz_2SSPTtoSSSB} is mapped to the Wen-plaquette model under $\mathbf{D}^{(2)}$; see the second row of table \ref{tab:GSD2dim}.
The trivial and non-trivial SSPTs have a single ground state. The SSSB phase has $2^{L_x+L_y-1}$ ground states and this can be seen by counting the number of independent stabilizer generators. The topological order (the Wen-plaquette model) has a  four or two-fold ground state degeneracy depending on whether it is an even by even lattice or not~\cite{wen2003quantum}. At the critical point between the trivial phase and SSSB phase (first-order transition), we expect $2^{L_x+L_y-1}+1$ ground states. The critical point between non-trivial SSPT and topological order is either gapless or has five or three ground states depending on the lattice size.  We summarize our discussions so far about ground state degeneracy of various models and their KW dual that are $\mathbb{Z}_2$ subsystem symmetric in table \ref{tab:GSD2dim} with two more examples.}
\begin{table}[h!]
    \centering
    \begin{tabular}{|c|c|c|c|}
    \hline
       $\rm H_1$ term & 
           $\rm H_2$ term
             & GSD of $\rm H_1$ & 
             GSD of $\rm H_2$ \\
        \hline\hline
        $\begin{array}{cc}
& X \\
&  
\end{array}$
            
& $\begin{array}{cc}
Z& Z \\
Z& Z 
\end{array}$ & 1 & $2^{L_x+L_y-1}$\\
\hline
        $\begin{array}{ccc}
 & Z & Z \\
Z& X & Z \\
Z& Z &
\end{array}$ & $\begin{array}{ccc}
 & Z & Y \\
 & Y & Z \\
 &  &
\end{array}$ & 1 & \begin{tabular}{@{}c@{}}
     4 for $L_x$, $L_y$ even,\\
     2 otherwise 
\end{tabular}\\
\hline
         $\begin{array}{ccc}
 & X & X \\
 &   &
\end{array}$ & $\begin{array}{ccc}
Z& I & Z \\
Z& I & Z
\end{array}$  & $2^{L_y}$ & \begin{tabular}{@{}c@{}}
    $2^{L_x+2L_y-2}$ \text{for} $L_x$ even, \\
    $2^{L_x+L_y-1}$ \text{for} $L_x$ odd
\end{tabular}\\ 
\hline
       $\begin{array}{ccc}
 &   & X \\
 & X &  \\
 &   &
\end{array}$  &  $\begin{array}{ccc}
 & Z & Z \\
Z&  & Z \\
Z& Z &
\end{array}$ & \begin{tabular}{@{}c@{}}
$2^{L}$ \\ for $L_x=L_y=L$
\end{tabular} 
& \begin{tabular}{@{}c@{}}
$2^{3L-2}$\\ for  $L_x=L_y=L$ 
\end{tabular}\\
\hline
    \end{tabular}
    \caption{Ground state degeneracy of various Hamiltonians with $\mathbb{Z}_2$ subsystem symmetry that we denote by $\rm H_1$ and $\rm H_2$. We omit the summation over lattice sites and overall negative signs in front of the Hamiltonian terms for simplicity. $\rm H_1$ and $\rm H_2$ are related by a Kramers-Wannier duality, i.e., $\rm \mathbf{D}^{(2)}\rm H_1=H_2\mathbf{D}^{(2)}$. }
    \label{tab:GSD2dim}
\end{table}
\red{For more general Hamiltonians that are invariant under both $\mathbb{Z}_2$ subsystem symmetry and non-invertible symmetry $\mathbf{D}^{(2)}$, counting the number of ground states requires a more careful analysis. 
This is because, first of all, the constraint about the number of ground states in the spirit of the LSM theorem is a question about the thermodynamic behavior of the lattice models that we start with. 
For lattice models with subsystem symmetries, the thermodynamic limit is very sensitive to the lattice system size. This is manifest in the extensive ground state degeneracy of these lattice models with large but finite system sizes. It would be interesting if one could define a notion of ground-state degeneracy in this setup as well as in higher dimensions and find an LSM-type constraint on the number of ground states. However, we leave this question for future exploration.}

\acknowledgments
The authors would like to thank Shu-Heng Shao, Yunqin Zheng and Mikhail Litvinov for useful discussions. APM and HS would like to thank Takuya Okuda for discussions on a project that uses similar ideas. This work was supported by the National Science Foundation under Award No. PHY 2310614. T.-C.W. also acknowledges the support by Stony Brook University's Center for Distributed Quantum Processing.

\bibliography{ref}
\appendix
\section{Fermionic dual}\label{sec:fermionicdual}
\subsection{Majorana plaquette model}
Let us consider the Majorana plaquette model in a 2D square lattice. At each vertex, we have two Majorana fermions $\gamma$ and $\gamma'$. 
The Hamiltonian is given by
\begin{align}
 H_{Maj-plaq}=\sum_{i,j}\left(\gamma'_{i,j}\gamma_{i+1,j}\gamma'_{i,j+1}\gamma_{i+1,j+1}+i\lambda \gamma_{i,j}\gamma'_{i,j}\right),
 \label{eq:Hmajplaq}
\end{align}
where $(i,j)$ denotes the $x$ and $y$ coordinates of the 2D square lattice. 
This model has the subsystem fermion parity symmetry along horizontal and vertical lines given by
\begin{align}
 S_j^x\equiv \prod_{i}\left(-i\gamma_{i,j}\gamma'_{i,j}\right),\qquad S_i^y\equiv\prod_j \left(-i\gamma_{i,j}\gamma'_{i,j}\right).
\end{align}
There are in total $2^{L_x+L_y}$ subsystem fermion parity symmetries. However, there is a global constraint
\begin{align}
 \prod_j S_j^x=\prod_i S_i^y.
\end{align}
Hence, in total, there are $2^{L_x+L_y-1}$ subsystem symmetries, which in turn agree with the number of subsystem symmetries for the plaquette Ising model. 
Now we discuss gauging the subsystem fermion parity. We consider the Jordan-Wigner transformation for subsystem symmetric fermionic models in 2D defined in Ref.~\cite{tantivasadakarn2020jordan,cao2022boson},
\begin{subequations}
\begin{align}
\gamma_{i,j}&\equiv\left(\prod_{i'=1}^{L_x}\prod_{j'=1}^{j-1}X_{i',j'}\right)\left(\prod_{i'=1}^{i-1}X_{i',j}\right)Z_{i,j},\\
\gamma_{i,j}'&\equiv -\left(\prod_{i'=1}^{L_x}\prod_{j'=1}^{j-1}X_{i',j'}\right)\left(\prod_{i'=1}^{i-1}X_{i',j}\right)Y_{i,j}.
\end{align}
\label{eq:JW}
\end{subequations}
Plugging the above relations in Eq.\eqref{eq:Hmajplaq}, we find the following Hamiltonian
\begin{widetext}
\begin{align}
\begin{split}
\tilde{H}_{plaq}&=-\sum_{i\neq L_x,j}Z_{i,j}Z_{i+1,j}Z_{i,j+1}Z_{i+1,j+1}-\sum_{j=1}^{L_y}\left(\prod_{i=1}^{L_x}X_{i,j}\right)\left(\prod_{i=1}^{L_x}X_{i,j+1}\right)Z_{L_x,j}Z_{1,j}Z_{L_x,j+1}Z_{1,j+1} -\lambda\sum_{i,j}X_{i,j}.
\end{split}
\label{eq:Hplaq}
\end{align}
\end{widetext}
This is the plaquette Ising model Hamiltonian with defects along the horizontal direction. The second term in Eq.~\eqref{eq:Hplaq} contains defects inserted at two consecutive horizontal lines. Note that the subsystem fermion parity maps to the subsystem line symmetry of the plaquette Ising model under the Jordan-Wigner transformation Eq.~\eqref{eq:JW},
\begin{align}
    S_j^x=\prod_{i=1}^{L_x}X_{i,j},\qquad S_i^y=\prod_{j=1}^{L_y}X_{i,j}.
\end{align}
To gauge the subsystem fermion parity, we need to sum over defect configurations(flipping the signs of terms in the second sum in Eq.~\eqref{eq:Hplaq}) only along the horizontal direction.

We perform a procedure similar to the procedure of gauging fermion parity in the free Majorana fermion model to obtain the transverse field Ising model in 1D~\cite{seiberg2023majorana}. 
First, let us define an extended Hilbert space.
\begin{align}
    \mathcal{H}\equiv\mathcal{H}_0\bigoplus_{\{i_1,...,i_k\}\neq \emptyset} \mathcal{H}_{i_1...i_k},\qquad \{i_1,...,i_k\}\subset \{1,...,L_y\}
\end{align}
where $\emptyset$ is the empty set and $i_1 < ... < i_k$. Now let us introduce a total ordering on the subsets of the set $\{1,2,...,L_y\}$. Suppose $A$ and $B$ are two subsets of $\{1,2,...,L_y\}$, then $A<B$ \text{ if } $|A|<|B|$\red{($|S|$ denote the cardinality of the set $S$)} or if $|A|=|B|$ then the least element in $A\cup B-A\cap B$ is contained in $A$.
With this ordering on the subsets, we define the Hamiltonian on the extended Hilbert space as
\begin{align}
    \begin{split}
      H&\equiv Diag(H_0,H_1,...,H_{L_y},...,H_{i_1...i_k},...,H_{1...L_y}),\\
      &\qquad \{i_1,...,i_k\}\subset\{1,...,L_y\}  
    \end{split}
    \label{eq:Hdiag}
\end{align}
where the entrees in the diagonal are ordered with respect to the subscript that is in one-to-one correspondence with the subsets of $\{1,2,...,L_y\}$. Hamiltonian $H_{i_1...i_k}$ denote the Hamiltonian Eq.~\eqref{eq:Hplaq} with defects inserted at rows labelled by $i_1$,...,$i_k$. 
Namely, the twisted Hamiltonian $H_{i_1...i_k}$ is defined as the Hamiltonian $\tilde{H}_{plaq}$ but with the product $S^x_j$ (\red{or} $S^x_{j+1}$) in the second term given a phase $(-1)$ when $j \in \{i_1,...,i_k\}$(\red{or $j+1 \in \{i_1,...,i_k\}$}). The untwisted Hamiltonian $H_0$ is the same as the Hamiltonian $\tilde{H}_{plaq}$ in Eq.~\eqref{eq:Hplaq} without any defects inserted. 
In total there are $2^{L_y}$ \red{isomorphic} copies of the Hilbert space \red{$\mathcal{H}_0$} in the total Hilbert space \red{$\mathcal{H}$} and $2^{L_y}$ Hamiltonians in the diagonal matrix that represent the Hamiltonian \red{$H$} in the total Hilbert space \red{$\mathcal{H}$}. 
The dimension of the extended Hilbert space $\mathcal{H}$ is $dim(\mathcal{H}_0)2^{L_y}$. 
We define the subsystem fermion parity on the total Hilbert space
\begin{align}
    (-1)^{\hat{\mathbf{F}}_i}\equiv Diag((-1)^{\hat{F}_i},...,(-1)^{\hat{F}_i}),
\end{align}
with $2^{L_y}$ copies of the subsystem fermion parity operator $(-1)^{\hat{F}_i}$ at $i^\text{th}$ row  on the diagonal. 
In the extended Hilbert space \red{$\mathcal{H}$}, the Hamiltonian $H$ has other $\mathbb{Z}_2$ symmetries which we denote by $\eta_i$. These are $\eta_i=Diag\left(a_0,a_1,...,a_{L_y},...,a_{i_1..i_k},...,a_{1...L_y}\right)$ with 
\begin{subequations}
\begin{align}
    a_{i_1...i_k}&=\begin{cases}
      -1 \quad\text{when}\quad i\in\{i_1,...,i_k\}\\
      1\quad\text{when}\quad i\notin\{i_1,...,i_k\}
    \end{cases}\\
    a_0&=1.
\end{align}
\end{subequations}
We perform a set of projections to preserve the dimension of the Hilbert space. The projections are
\begin{align}
    \eta_i(-1)^{\hat{\mathbf{F}}_i}=I,
    \label{eq:projections}
\end{align}
where $I$ is the $2^{L_y}\times 2^{L_y}$ identity matrix.
Then $\text{dim}(\mathcal{H}\rvert_{\eta_i(-1)^{\hat{\mathbf{F}}_i}=1})=\text{dim}(\mathcal{H}_0)$ as we want. Now we write down a representation of the Pauli operators in the extended Hilbert space. We take the Pauli $X$ to be the diagonal matrix
\begin{align}
    X\equiv\begin{pmatrix}
        X &        & \\
          & \ddots & \\
          &        & X
    \end{pmatrix}.
    \label{eq:newpauliX}
\end{align}
However, we cannot take the Pauli $Z$ and $Y$ to be diagonal since they wouldn't commute with the projection.
\red{We define the new Pauli $Z$ and $Y$ operators that commute with the projection in terms of their matrix elements. This definition depend on the vertical position of the operators on the square lattice. Let $S\subset\{1,...,L_y\}$
\begin{subequations}
\begin{align}
   \left( Z_{-,j}\right)_{S\cup\{j\},S}= \left( Z_{-,j}\right)_{S,S\cup\{j\}}=Z_{-,j}\, ,\\
   \left( Y_{-,j}\right)_{S\cup\{j\},S}=\left( Y_{-,j}\right)_{S,S\cup\{j\}}=Y_{-,j}\, ,
\end{align}
\label{eq:newpauliZandY}
\end{subequations}
where the index of subsets is numbered, according to the total ordering we defined above, starting from one. The Pauli $Z$ and $Y$ operators on the R.H.S. is acting on the Hilbert space $\mathcal{H}_{S}$ or $\mathcal{H}_{S\cup\{j\}}$. It is an easy exercise to see that above defined operators indeed satisfy the Pauli algebra.}
With these definitions of Pauli operators, the Hamiltonian $H$ in Eq.~\eqref{eq:Hdiag} is
\begin{align}
    H_{plaq}=-\sum_{i,j}Z_{i,j}Z_{i+1,j}Z_{i,j+1}Z_{i+1,j+1}-\lambda\sum_{i,j}X_{i,j}.
\end{align}
\red{This is the plaquette Ising model Hamiltonian.}
\subsubsection{Anomalous symmetry}
As we discussed before, the Hamiltonian Eq.~\eqref{eq:Hmajplaq} has subsystem fermion parity symmetry. At $\lambda=1$, in addition to this symmetry, there is an exchange symmetry that exchanges the plaquette term with the onsite fermion parity term,
\begin{align}
    E: \begin{cases}
        i\gamma_{i,j}\gamma'_{i,j}\rightarrow \gamma'_{i,j}\gamma_{i+1,j}\gamma'_{i,j+1}\gamma_{i+1,j+1}\\
        \gamma'_{i,j}\gamma_{i+1,j}\gamma'_{i,j+1}\gamma_{i+1,j+1}\rightarrow i\gamma_{i+1,j+1}\gamma'_{i+1,j+1}.
    \end{cases} 
\end{align}
This symmetry does not commute with the subsystem fermion parity and hence is anomalous. After gauging the subsystem fermion parity, the exchange symmetry $E$ gives rise to the non-invertible symmetry $\mathbf{D}^{(2)}$ at $\lambda=1$.

\subsection{Majorana hypercubic models}
The same gauging procedure can be carried out in higher dimensions. We start with hypercubic Majorana models and gauge the subsystem fermion parity to obtain hypercubic Ising models. Let us consider $d$ dimensions and the following fermionic Hamiltonian,
\begin{widetext}
    \begin{align}
\begin{split}
    H_{Maj-HC}&=\sum_{i_1,...,i_d}\left(-\prod_{j_2,...,j_d=0}^1\gamma'_{i_1,i_2+j_2,...,i_d+j_d}\gamma_{i_1+1,i_2+j_2,...,i_d+j_d}+i\lambda\gamma_{i_1,...,i_d}\gamma'_{i_1,...,i_d}\right).
    \label{eq:MajHC}
\end{split}
\end{align}
This model has subsystem fermion parity along lines in any of the $d$ directions. 
\begin{align}
    S^{x_l}_{i_1,...\hat{i}_l,...,i_d}\equiv \prod_{i_l}\left(-i\gamma_{i_1,...,i_d}\gamma'_{i_1,...,i_d}\right).
\end{align}
We consider a generalization of the Jordan-Wigner transformation for subsystem symmetric fermionic models in higher dimensions,
\begin{subequations}
    \begin{align}
    \begin{split}
        \gamma_{i_1...i_d}&\equiv\left(\prod_{i_1'=1}^{L_{x_{1}}}\prod_{i_2'=1}^{L_{x_{2}}}...\prod_{i_d'=1}^{i_d-1}X_{i_1',...,i_d'}\right)\left(\prod_{i_1'=1}^{L_{x_{1}}}\prod_{i_2'=1}^{L_{x_{2}}}...\prod_{i_{d-1}'=1}^{i_{d-1}-1}X_{i_1',...,i_{d-1}',i_d}\right)... \left(\prod_{i_1'=1}^{i_1-1}X_{i_1',i_2,...,i_d}\right)Z_{i_1,...,i_d},
    \end{split}
    \end{align}
    \begin{align}
    \begin{split}
        \gamma'_{i_1...i_d}&\equiv -\left(\prod_{i_1'=1}^{L_{x_{1}}}\prod_{i_2'=1}^{L_{x_{2}}}...\prod_{i_d'=1}^{i_d-1}X_{i_1',...,i_d'}\right)\left(\prod_{i_1'=1}^{L_{x_{1}}}\prod_{i_2'=1}^{L_{x_{2}}}...\prod_{i_{d-1}'=1}^{i_{d-1}-1}X_{i_1',...,i_{d-1}',i_d}\right) ... \left(\prod_{i_1'=1}^{i_1-1}X_{i_1',i_2,...,i_d}\right)Y_{i_1,...,i_d}.
    \end{split}
\end{align}
\end{subequations}
We plug this transformation into the Hamiltonian Eq.~\eqref{eq:MajHC} and obtain
\begin{align}
\begin{split}
    \tilde{H}_{HC}&=-\sum_{\substack{i_1\neq L_{x_{1}}\\,i_2,...,i_d}}\prod_{j_2,...,j_d=0}^1Z_{i_1,i_2+j_2,...,i_d+j_d} Z_{i_1+1,i_2+j_2,...,i_d+j_d}\\
    &\hspace{1cm}-\sum_{i_2,...,i_d}\prod_{j_2,...j_d=0}^1\prod_{i_1=1}^{L_{x_{1}}}X_{i_1,i_2+j_2,...,i_d+j_d}\prod_{j_2,...,j_d=0}^1Z_{L_{x_{1}},i_2+j_2,...,i_d+j_d} Z_{1,i_2+j_2,...,i_d+j_d}-\lambda\sum_{i_1,...,i_d}X_{i_1,...,i_d}.
\end{split}
\label{eq:HHC}
\end{align}
\end{widetext}
This is the hypercubic Ising model Hamiltonian with defects inserted along $x_1$ direction. The second term in Eq.~\eqref{eq:HHC} contains the defect lines. To gauge the subsystem fermion parity, we sum over defect configurations in the $x_1$ direction. Define an extended Hilbert space
\begin{align}
    \begin{split}
        \mathcal{H}&\equiv \mathcal{H}_0\bigoplus_{\{i_1,...,i_k\}\neq \emptyset}\mathcal{H}_{i_1...i_k},\qquad
     i_j\in\{1,...,L_{x_2}...L_{x_d}\},\\
     &\qquad\{i_1,...,i_k\}\subset \{1,...,L_{x_2}...L_{x_d}\}.
    \end{split}
\end{align}
We choose an ordering on the subsystem fermion parity lines in the $x_1$ direction. Any line in the $x_1$ direction is specified by the $d-1$ of the remaining coordinates $x_2$,...,$x_d$. Then we choose an ordering on the remaining coordinates $(x_2,...,x_d)<(x_2',...,x_d')$ if $x_i<x_i'$ provided $x_j=x_j'$ for all $j<i$, i.e., ordering is chosen based on the first coordinate (reading from left) for which the two tuples disagree. Then these lines are enumerated from 1 to $L_{x_2}...L_{x_d}$. Similar to the two-dimensional case we define a total ordering on the subsets of $ \{1,...,L_{x_2}...L_{x_d}\}$ with the definition of the total ordering exactly as in the case of two dimensions. With respect to this ordering, we define the Hamiltonian on the extended Hilbert space
\begin{align}
    \begin{split}
       H&\equiv Diag\left(H_0,H_1,...,H_{i_1...i_k},...,H_{1...(L_{x_2}...L_{x_d})}\right),\\
       &
    \qquad\qquad\quad i_j\in\{1,...,L_{x_2}...L_{x_d}\},\\
    &\qquad\quad\{i_1,...,i_k\}\subset\{1,...,L_{x_2}...L_{x_d}\}, 
    \end{split}
    \label{eq:totalH}
\end{align}
where the entrees in the diagonal are ordered with respect to the subscript that is in one-to-one correspondence with the subsets of $\{1,2,...,L_{x_2}...L_{x_d}\}$. Hamiltonian $H_{i_1...i_k}$ denotes the Hamiltonian Eq.~\eqref{eq:HHC} with defects inserted on lines labelled by $i_1$,$i_2$,... up to $i_k$. Namely, the twisted Hamiltonian $H_{i_1...i_k}$ is defined as the Hamiltonian $\tilde{H}_{HC}$ but with the subsystem fermion parity lines enumerated as $j$ (according to our definition of enumeration) in the second term given a phase $(-1)$ when $j \in \{i_1,...,i_k\}$. The untwisted Hamiltonian $H_0$ is the same as the Hamiltonian $\tilde{H}_{HC}$ in Eq.~\eqref{eq:Hplaq} without any defects inserted. In total, there are $2^{L_{x_2}...L_{x_d}}$ \red{isomorphic} copies of the Hilbert space \red{$\mathcal{H}_0$} in the total Hilbert space \red{$\mathcal{H}$}. The rest of the analysis is a straightforward extension of the one we performed for the Majorana plaquette model in the previous subsection with $i$ taking values in $\{1,...,L_{x_2}...L_{x_d}\}$. After the projections Eq.~\eqref{eq:projections}, we obtain the total Hamiltonian Eq.~\eqref{eq:totalH} with the new Pauli $X$ and $Z$ operators similar as in Eqs.~\eqref{eq:newpauliX},\eqref{eq:newpauliZandY}
\begin{widetext}
    \begin{align}
\begin{split}
    H_{HC}&=-\sum_{i_1\neq L_{x_{1}},i_2,...,i_d}\prod_{j_2,...,j_d}Z_{i_1,i_2+j_2,...,i_d+j_d} Z_{i_1+1,i_2+j_2,...,i_d+j_d}-\lambda\sum_{i_1,...,i_d}X_{i_1,...,i_d}.
\end{split}
\end{align}
\red{This is the Hamiltonian for hypercubic Ising model.}
\end{widetext}
\section{KT transformation as $TST$} \label{sec:TST}
 In this section, we look at an alternate way of writing down the KT transformation. 
 \subsection{One dimension}
 \red{We consider a ring with $2M$ sites}. Let us consider the operator 
 \begin{align}
     \overline{\text{KT}}\equiv T\mathbf{D}_{even}\mathbf{D}_{odd}T .
     \label{eq:1DKTbar}
 \end{align}
 After simplifying Eq.~\eqref{eq:1DKTbar}, $\overline{\text{KT}}$ can be written explicitly as
\begin{align}
    \overline{\text{KT}}=\overline{\mathbf{D}}_{even}\overline{\mathbf{D}}_{odd},
\end{align}
where 
\begin{subequations}
    \begin{align}
    \begin{split}
        \overline{\mathbf{D}}_{even}&\equiv\left(\prod_{k=1}^{\red{M}-1}e^{i\frac{\pi}{4}Z_{2k-1}X_{2k}Z_{2k+1}}e^{i\frac{\pi}{4}Z_{2k}Z_{2k+2}}\right) \\
        &\quad\times e^{i\frac{\pi}{4}Z_{2\red{M}-1}X_{2\red{M}}Z_1}\frac{\left(1+\eta_{even}\right)}{2} ,
    \end{split}\\
    \begin{split}
        \overline{\mathbf{D}}_{odd}&\equiv\left(\prod_{k=1}^{\red{M}-1}e^{i\frac{\pi}{4}Z_{2k-2}X_{2k-1}Z_{2k}}e^{i\frac{\pi}{4}Z_{2k-1}Z_{2k+1}}\right)\\
        &\times e^{i\frac{\pi}{4}Z_{2\red{M}-2}X_{2\red{M}-1}Z_{2\red{M}}}\frac{\left(1+\eta_{odd}\right)}{2} .
    \end{split}
\end{align}
\label{eq:KTbarevenodd}
\end{subequations}
Note that the pattern in the product of operators in Eq.~\eqref{eq:KTbarevenodd} takes an interesting form; alternating terms of the form $ZXZ$ and $ZZ$ in the exponents. In fact, this pattern gives rise to the desired Kennedy-Tasaki transformation. Explicitly, the action of our $\overline{\text{KT}}$ transformation is given by
\begin{subequations}
    \begin{align}
    \overline{\text{KT}}X_i&=X_{i+1}\overline{\text{KT}} , \\
    \overline{\text{KT}}Z_{i-1}X_iZ_{i+1}&=Z_iZ_{i+2}\overline{\text{KT}} . 
\end{align}
\label{eq:KTbartrans1D}
\end{subequations}
Hence, $\overline{\text{KT}}$ maps the cluster Hamiltonian to two (decoupled) copies of Ising models in the two sublattices. On a single $Z$ operator, the action of $\overline{\text{KT}}$ is given by 
\begin{subequations}
\begin{align}
    \overline{\text{KT}}Z_{2i+1}&=Z_{2\red{M}}Y_1\prod_{k=1}^{i}X_{2k+1}Z_{2i+2}\overline{\text{KT}}', \\ 
    \overline{\text{KT}}Z_{2i}&=Z_1Y_2\prod_{k=2}^{i}X_{2k}Z_{2i+1}\overline{\text{KT}}'' ,
\end{align}
\end{subequations}
where $\overline{\text{KT}}'$ and $\overline{\text{KT}}''$ are defined as
\begin{subequations}
\begin{align}
    \overline{\text{KT}}'\equiv T\mathbf{D}_{even}\mathbf{D}_{odd}'T,\\
    \overline{\text{KT}}''\equiv T\mathbf{D}_{even}'\mathbf{D}_{odd}T,
\end{align}
\end{subequations}
with 
\small{
\begin{subequations}
\begin{align}
    \mathbf{D}_{odd}'&\equiv\Big(\prod_{k=1}^{\red{M}-1} e^{i\frac{\pi}{4}X_{2k-1}}e^{i\frac{\pi}{4}Z_{2k-1}Z_{2k+1}}\Big) e^{i\frac{\pi}{4}X_{2\red{M}-1}}\frac{(1-\eta_{odd})}{2},\\
    \mathbf{D}_{even}'&\equiv\Big(\prod_{k=1}^{N-1} e^{i\frac{\pi}{4}X_{2k}}e^{i\frac{\pi}{4}Z_{2k}Z_{2k+2}}\Big) e^{i\frac{\pi}{4}X_{2\red{M}}}\frac{(1-\eta_{even})}{2}.
\end{align}
\end{subequations}}
\normalsize
Similarly,
\begin{subequations}
\begin{align}
    \overline{\text{KT}}^{\dagger}Z_{2i}=-Z_{2i-1}\prod_{k=i}^{\red{M}-1}X_{2k}Y_{2\red{M}}Z_1(\overline{\text{KT}}'')^{\dagger} ,\\
    \overline{\text{KT}}^{\dagger}Z_{2i+1}=-Z_{2i}\prod_{k=i}^{\red{M}-2}X_{2k+1}Y_{2\red{M}-1}Z_{2\red{M}}(\overline{\text{KT}}')^{\dagger}.
\end{align}
\end{subequations}
Composition of $\overline{\text{KT}}$ with itself gives
\begin{align}
    \overline{\text{KT}}^2=\red{e^{i\pi M}}\mathbf{T}_2\frac{(1+\eta_{even})}{2}\frac{(1+\eta_{odd})}{2},
\end{align}
where $\mathbf{T}_2$ is a translation by two lattice sites. The projection factor implies that $\overline{\text{KT}}$ is non-invertible.
\subsection{Two dimensions}
We generalize the $TST$ version of KT transformation to two dimensions.
 \begin{align}
     \overline{\text{KT}}^{(2)}\equiv T^{(2)}\mathbf{D}_r^{(2)}\mathbf{D}_b^{(2)}T^{(2)}.
 \end{align}
 Explicitly $\overline{\text{KT}}^{(2)}$ transformation is given by
 \begin{widetext}
 \begin{subequations}
\begin{align}
    &\overline{\text{KT}}^{(2)}X^r_{i,j}=X^b_{i+\frac{1}{2},j+\frac{1}{2}}\overline{\text{KT}}^{(2)}\,\\
    \begin{split}
        &\overline{\text{KT}}^{(2)} X^b_{i+\frac{1}{2},j+\frac{1}{2}}Z^r_{i,j}Z^r_{i+1,j}Z^r_{i,j+1}Z^r_{i+1,j+1}=Z^b_{i+\frac{1}{2},j+\frac{1}{2}}Z^b_{i+1+\frac{1}{2},j+\frac{1}{2}}Z^b_{i+\frac{1}{2},j+1+\frac{1}{2}}Z^b_{i+1+\frac{1}{2},j+1+\frac{1}{2}}\overline{\text{KT}}^{(2)}
    \end{split}\,
\end{align}
 \end{subequations}
and similar equations for $r\leftrightarrow b$. Hence $\overline{\text{KT}}^{(2)}$ maps $\mathbb{Z}_2\times\mathbb{Z}_2$ cluster SSPT Hamiltonian Eq.~\eqref{eq:2dcluster} to two copies of the plaquette Ising model, which is in the SSSB phase. $\overline{\text{KT}}^{(2)}$ also maps the $\mathbb{Z}_2$ SSPT Hamiltonian Eq.~\eqref{eq:sixZSSPT} to a single copy of double plaquette-Ising model Hamiltonian in Eq.~\eqref{eq:sixZIsing} that is in the SSSB phase.

Action of $\overline{\text{KT}}^{(2)}$ on a single $Z$ operator is 
\begin{align}
    \begin{split}
        \overline{\text{KT}}^{(2)}Z^r_{i,j}&=(-i)Z_{i,1}^rZ^r_{i+1,1}Z^b_{\frac{1}{2},j+\frac{1}{2}}Z^b_{i+\frac{1}{2},j+\frac{1}{2}}Z^b_{\frac{1}{2},\frac{1}{2}}Z^b_{i+\frac{1}{2},\frac{1}{2}} Y^r_{1,1}\prod_{k=2}^jX^r_{k,1}\prod_{m=2}^j\left(Y_{1,m}^r\prod_{l=2}^i X_{l,k}^r\right)\overline{\text{KT}}^{(2)'},
    \end{split}
    \label{eq:KT2singleZ}
\end{align}
where 
\begin{align}
    \overline{\text{KT}}^{(2)'}\equiv T^{(2)}\bm \textbf{D}_r^{(2)'}\textbf{D}_b^{(2)} T^{(2)},
\end{align}
with 
\begin{align}
    \textbf{D}_r^{(2)'}&\equiv\mathbf{\tilde{P}}^{\red{(2)}'}_{\red{r}}\tilde{\mathbf{D}}_x\red{\mathbf{H}^{\otimes(2)}_{r}}\tilde{\mathbf{D}}_y\mathbf{\tilde{P}}^{\red{(2)}}_{\red{r}},
\end{align}
and
\begin{subequations}
\begin{align}
    \begin{split}
        \mathbf{\tilde{P}}^{\red{(2)}'}_{\red{r}}&\equiv\prod_{l=1}^j\frac{\left(1-\eta^x_l\right)}{2}\prod_{l=j+1}^{L_y}\frac{\left(1+\eta^x_l\right)}{2}\prod_{k\neq 1,i,i+1}\frac{\left(1+\eta^y_k\right)}{2}\frac{(1-\eta^y_i)}{2}\frac{(1-\eta^y_{i+1})}{2}\frac{(1+(-1)^j\eta^y_1)}{2},
    \end{split}\\
    \mathbf{\tilde{P}}^{\red{(2)}}_{\red{r}}&\equiv\prod_{l\neq j}\frac{(1+\eta^x_l)}{2}\prod_{k\neq i}\frac{(1+\eta^y_k)}{2}\frac{(1-\eta^x_j)}{2}\frac{(1-\eta^y_i)}{2},
\end{align}
\end{subequations}
 \end{widetext}
and similar equation for $r\leftrightarrow b$. 

Furthermore, our mapping of the Pauli $X$ operator is shifted diagonally to the other sublattice. We also find the composition of two $\overline{\text{KT}}^{(2)}$ transformations gives
\begin{align}
    (\overline{\text{KT}}^{(2)})^2\propto\,\mathbf{T}_{\red{(1,1),}r}\mathbf{T}_{\red{(1,1),}b}\mathbf{P}^{\red{(2)}}_r\mathbf{P}^{\red{(2)}}_b,
\end{align}
where $\mathbf{T}_{\red{(1,1),}r}$ and $\mathbf{T}_{\red{(1,1),}b}$ are diagonal translations on the red and blue sublattices, respectively. $\mathbf{P}^{\red{(2)}}_r$ and $\mathbf{P}^{\red{(2)}}_b$ are the respective projections onto the Hilbert spaces of the red and blue sublattices, which are separately subsystem symmetric.

\begin{figure}[h!]
    \centering
    \includegraphics[width=0.3\textwidth]{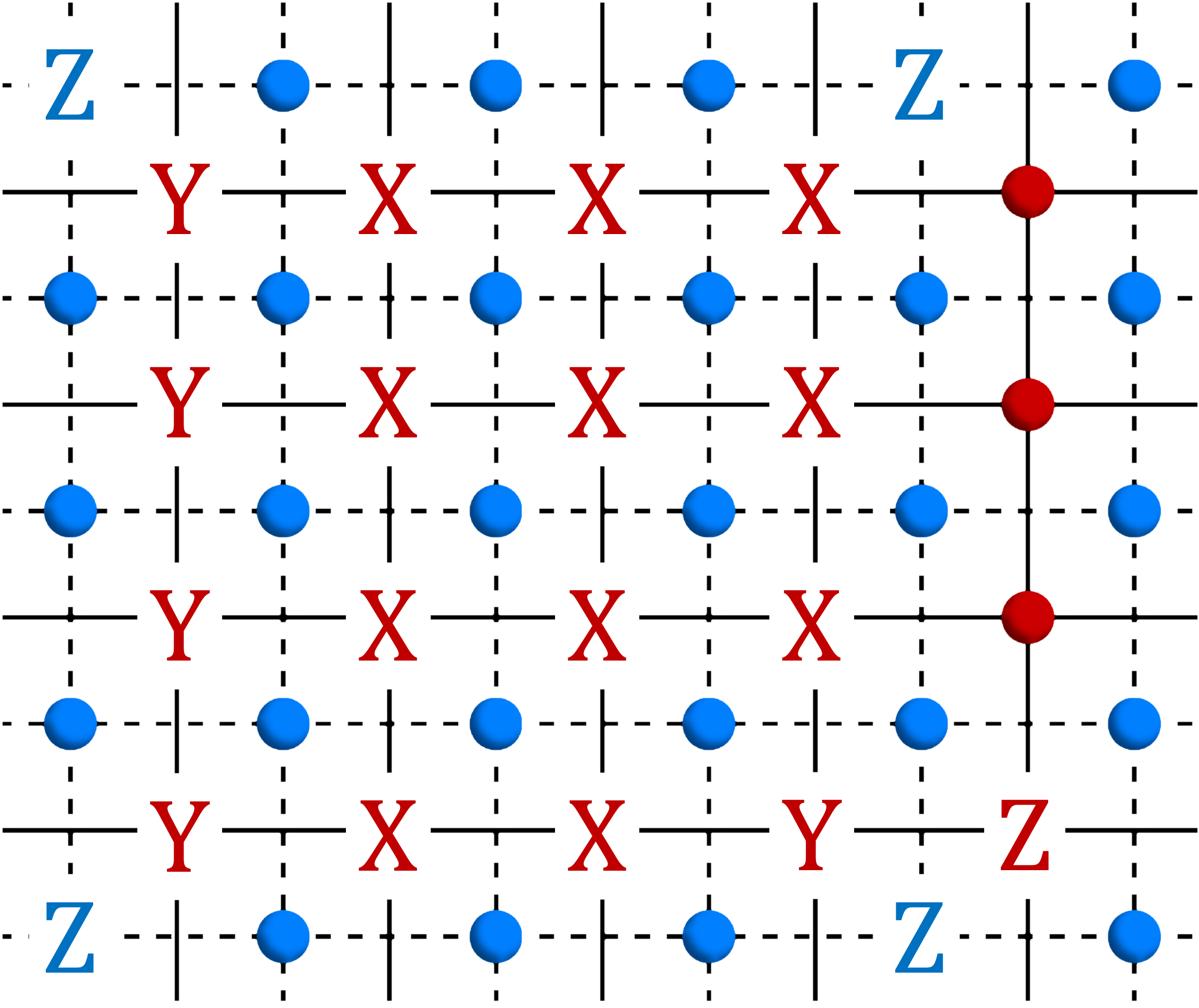}
    \caption{$\overline{\text{KT}}^{(2)}$ transformation acting on a single $Z$ operator at the red site is given by a membrane operator according to Eq.~\eqref{eq:KT2singleZ}.}
    \label{fig:KTZtransquare}
\end{figure}
    
 \subsection{Higher dimensions}
 Explicitly $TST$ version of KT transformation is
 \begin{subequations}
\begin{align}
    \overline{\text{KT}}^{(d)}X_{v^r}&=X_{v^b}\overline{\text{KT}}^{(d)},\\
    \overline{\text{KT}}^{(d)}X_{v^b}\prod_{v^r\in \partial c^r}Z_{v^r}&=\prod_{v^b\in\partial c^b}Z_{v^b}\overline{\text{KT}}^{(d)},
\end{align}
 \end{subequations}
where the terms in the L.H.S. and R.H.S. are related by diagonal translation by half unit. By $r\leftrightarrow b$ we get another similar set of equations. Composition of two $\overline{\text{KT}}^{(d)}$ transformations is
\begin{align}
    (\overline{\text{KT}}^{(d)})^{2}\red{\propto}\,\mathbf{T}_{\red{(1,...,1),}r}\mathbf{T}_{\red{(1,...,1),}b}\mathbf{P}_r^{(d)}\mathbf{P}_b^{(d)} ,
\end{align}
where $\red{\mathbf{T}_{\red{(1,...,1),}r}}$ and $\red{\mathbf{T}_{\red{(1,...,1),}b}}$ are diagonal translations on the red and blue sublattices. $\mathbf{P}_r^{(d)}$ and $\mathbf{P}_b^{(d)}$ are projections onto subsystem symmetric subspace of the Hilbert spaces of red and blue sublattices.

\section{$\mathbb{Z}_N$ generalization of hypercubic Ising models}\label{sec:Z_Nhigherdimmodels}
\subsection{$\mathbb{Z}_N$ plaquette transverse field clock model}
As before, we consider a 2D square lattice with $L_x$ sites in the $x$-direction and $L_y$ sites in the $y$-direction with the periodic boundary condition. The Hamiltonian for the $\mathbb{Z}_N$ generalization to the plaquette clock model, which we call the transverse-field plaquette clock model, is
    \begin{align}
    \begin{split}
    H_{TFPC}&=-\sum_{i,j}\left(Z_{i,j}Z^{\dagger}_{i+1,j}Z^{\dagger}_{i,j+1}Z_{i+1,j+1}+\text{h.c}\right.\\
    &\left.\hspace{3cm}+\lambda(X_{i,j}+X_{i,j}^{\dagger})\right).
     \end{split}
\end{align}
This model has $\mathbb{Z}_N$ subsystem symmetry along horizontal rows and vertical columns. Let us denote them by
\begin{align}
    \eta^x_j\equiv \prod_{i}X_{i,j},\quad \eta^y_i\equiv \prod_{j}X_{i,j}.
\end{align}
They satisfy $(\eta_j^x)^N=1$ and $(\eta_i^y)^N=1$.
At $\lambda=1$, there is an extra symmetry that exchanges the clock term and the transverse-field term. We give an explicit expression for this symmetry operator as a generalization of Eq.~\eqref{eq:2DDsym}. Let us define
      \begin{align}
     \Tilde{\mathbf{D}}_{(N),x}\equiv \prod_{j=1}^{L_y}\red{\Tilde{\mathbf{D}}_{(N),j}^x},\qquad
     \Tilde{\mathbf{D}}_{(N),y}\equiv \prod_{j=1}^{L_x}\red{\Tilde{\mathbf{D}}_{(N),j}^y},
 \end{align}
 where $\red{\Tilde{\mathbf{D}}_{(N),j}^x}$ represent the operator $\red{\Tilde{\mathbf{D}}_{(N)}}$ for a fixed horizontal row. Similarly $\red{\Tilde{\mathbf{D}}_{(N),j}^y}$ represent the operator $\red{\Tilde{\mathbf{D}}_{(N)}}$ for a fixed vertical column. We define the following operator, a quantum Fourier transform that is a generalization of the Hadamard gate to qudit d.o.f.,
 \begin{align}
     \red{\mathbf{H}}_{(N)}\equiv \frac{1}{\sqrt{N}}\sum_{x=0}^{N-1}\sum_{y=0}^{N-1}e^{\frac{2\pi i x y}{N}}\ket{x}\bra{y}.
 \end{align}
 This operator implements the following relation 
     \begin{align}
     \red{\mathbf{H}}_{(N)}Z=X^{\dagger}\red{\mathbf{H}}_{(N)},\qquad
     \red{\mathbf{H}}_{(N)}X=Z\red{\mathbf{H}}_{(N)}.
 \end{align}
 Note that the operator $\mathbf{H}_{(N)}$ for $N=2$ is the same as the Hadamard transformation.
 We define the operator
 \begin{widetext}
 \begin{align}
     \Tilde{\mathbf{D}}_{(N)}^{\red{(2)}}\equiv\Tilde{\mathbf{D}}_{(N),x}\red{\mathbf{H}^{\otimes(2)}_{(N)}}\Tilde{\mathbf{D}}_{(N),y},\quad \mathbf{P}_{(N)}^{\red{(2)}}\equiv\prod_{j=1}^{L_y}\frac{(1+\eta^x_j+(\eta^x_j)^2+...+(\eta^x_j)^{N-1})}{N}\prod_{i=1}^{L_x}\frac{(1+\eta^y_i+(\eta^y_i)^2+...+(\eta^y_i)^{N-1})}{N}.
 \end{align}
 \end{widetext}
 \red{where $\mathbf{H}^{\otimes(2)}_{(N)}$ represent the tensor product of $\mathbf{H}_{(N)}$ on all sites of the two-dimensional lattice.}
The non-invertible KW duality operator we define is $\mathbf{D}^{\red{(2)}}_{(N)}=\mathbf{P}_{(N)}^{\red{(2)}}\Tilde{\mathbf{D}}_{(N)}^{\red{(2)}}\mathbf{P}_{(N)}^{\red{(2)}}$. The following relations are satisfied by $\mathbf{D}_{(N)}^{\red{(2)}}$,
 \begin{subequations}
     \begin{align}
     &\mathbf{D}_{(N)}^{\red{(2)}}X_{i,j}=Z_{i,j}^{\red{\dagger}}Z_{i+1,j}Z_{i,j+1}Z_{i+1,j+1}^{\red{\dagger}}\mathbf{D}_{(N)}^{\red{(2)}},\\
     &\mathbf{D}_{(N)}^{\red{(2)}}Z_{i,j}^{\red{\dagger}}Z_{i+1,j}Z_{i,j+1}Z_{i+1,j+1}^{\red{\dagger}}=X^{\dagger}_{i+1,j+1}\mathbf{D}_{(N)}^{\red{(2)}}.
 \end{align}
 \end{subequations}
 The operator $\mathbf{D}_{(N)}^{\red{(2)}}$ commutes with the Hamiltonian at $\lambda=1$. We also have $(\mathbf{D}_{(N)}^{\red{(2)}})^2\red{\propto}\,\mathbf{C}\mathbf{T}_{\red{(1,1)}}\mathbf{P}_{(N)}^{\red{(2)}}$ where $\mathbf{T}_{\red{(1,1)}}$ is the diagonal translation operator which send $(i,j)$ to $(i+1,j+1)$ and $\mathbf{C}$ is the conjugation operator with action $\red{\mathbf{C}}X=X^{\dagger}\red{\mathbf{C}}$ and $\red{\mathbf{C}}Z=Z^{\dagger}\red{\mathbf{C}}$.
 \subsection{higher-dimensional hypercubic clock models}
 We generalize the discussion for the hypercubic clock model with a transverse field to all spatial dimensions. The Hamiltonian for the transverse field hypercubic clock model is given by
 \begin{align}
    \bm H_{\text{TFHCC}}=-\sum_{c\in\Delta_c}\left(\prod_{v\in\partial c}Z_{v}+\text{h.c}\right)-\lambda\sum_{v\in\Delta_v}(X_v+X_v^{\dagger}).
\end{align}
 The non-invertible symmetry operator (at $\lambda=1$) is generalized as
 \begin{align}
     \mathbf{D}_{(N)}^{(d)}\equiv \mathbf{P}_{(N)}^{\red{(d)}}\Tilde{\mathbf{D}}_{(N),d}\red{\mathbf{H}^{\otimes(d)}_{(N)}}\Tilde{\mathbf{D}}_{(N),d-1}\red{\mathbf{H}^{\otimes(d)}_{(N)}}...\Tilde{\mathbf{D}}_{(N),1}\mathbf{P}_{(N)}^{\red{(d)}},
 \end{align}
 where each of the $\Tilde{\mathbf{D}}_{(N),i}$ are string operators like Eq.~\eqref{eq:Dtilde} on straight line along $i^\text{th}$ direction\red{, $\red{\mathbf{H}^{\otimes(d)}_{(N)}}$ represent the tensor product of $\mathbf{H}_{(N)}$ on all sites of the $d$ dimensional lattice} and $\mathbf{P}_{(N)}^{\red{(d)}}$ is a projection onto subsystem straight line like symmetries. The operator $\mathbf{D}_{(N)}^{(d)}$ commutes with the Hamiltonian at $\lambda=1$. \red{Up to an overall normalization} for $\mathbf{D}_{(N)}^{(d)}$, acting the operator twice will generate a diagonal unit shift and a conjugation in d dimensions, i.e, $(\mathbf{D}_{(N)}^{(d)})^2\red{\propto}\,\mathbf{C}\red{\mathbf{T}_{(1,...,1)}}\mathbf{P}_{(N)}^{\red{(d)}}$ where $\red{\mathbf{T}_{(1,...,1)}}$ is a diagonal translation on the hypercube and $\mathbf{C}$ is a conjugation.
 \section{$\mathbf{Z}_N$ generalization of KT transformation}\label{sec:Z_NgeneralKT}
 \subsection{One dimension}
 Let us consider the $\mathbb{Z}_N\times\mathbb{Z}_N$ SPTs in one dimension. According to the classification of SPTs in Ref.~\cite{chen2012symmetry}, we have $H^2(\mathbb{Z}_N\times\mathbb{Z}_N,U(1))=\mathbb{Z}_N$. Hence, there are non-trivial $\mathbb{Z}_N\times\mathbb{Z}_N$ SPTs in one dimension. Let us consider the following non-trivial SPT Hamiltonian on a ring with $2L$ sites~\cite{tsui2017phase},
 \begin{widetext}
 \begin{eqnarray}
 \begin{split}
         H^{l,N}_{1DSPT}&=-\sum_{i=1}^L\left(\left(Z_{2i-2}^{\dagger}\right)^lX_{2i-1}Z_{2i}^l+Z_{2i-1}^lX_{2i}\left(Z_{2i+1}^{\dagger}\right)^l+\text{h.c}\right).
         \end{split}
 \end{eqnarray}
  \end{widetext}
 This Hamiltonian has $\mathbb{Z}_N\times\mathbb{Z}_N$ symmetry generated by $\eta_1=\prod_{i=1}^LX_{2i-1}$ and $\eta_2=\prod_{i=1}^LX_{2i}$. This Hamiltonian can be obtained from the trivial Hamiltonian 
 \begin{align}
     H_{\text{triv}}=-\sum_{i=1}^{2L}\left(X_i+X_i^{\dagger}\right),
 \end{align}
 using $\mathbb{Z}_N$ cluster entangler. To define the $\mathbb{Z}_N$ cluster entangler, first, we define the generalization of the controlled-Z
 gate to qudits.
 \begin{align}
     \text{CZ}^{k}_{c,t}\equiv \sum_{n=0}^{N-1}\ket{n}_c\bra{n}\otimes Z_t^{kn},\quad  \text{CZ}^{\dagger}_{c,t}\equiv \text{CZ}^{N-1}_{c,t}.
 \end{align}
 Then the cluster entangler is defined as
 \begin{align}
     T\equiv \prod_{i=1}^{L} \text{CZ}_{2i,2i-1}\text{CZ}^{\dagger}_{2i,2i+1},\qquad T^N=1.
 \end{align}
 Note that $H_{1DSPT}^{l,N}$ is obtained from $H_{\text{triv}}$ using $l^{th}$ power of cluster entangler $T$.
 Now we define $\text{KT}_{l,N}$ that maps $\mathbb{Z}_N\times\mathbb{Z}_N$ SPT to two copies of clock models. 
 \begin{align}
     \text{KT}_{l,N}\equiv (T^{\dagger})^l\mathbf{D}_{(N),\text{odd}}\mathbf{D}_{(N),\text{even}}(T^{\dagger})^l,
 \end{align}
 where $\mathbf{D}_{(N),odd}$ and $\mathbf{D}_{(N),even}$ represent the KW duality operator $\mathbf{D}_{(N)}$ on odd and even sites.
 It satisfies the following properties
 \begin{subequations}
     \begin{align}
     \text{KT}_{l,N} Z_{2i-1}^lX_{2i}(Z_{2i+1}^{\dagger})^l&=Z_{2i}Z_{2i+2}^{\red{\dagger}}\text{KT}_{l,N},\\
     \text{KT}_{l,N} (Z_{2i}^{\dagger})^lX_{2i+1}Z_{2i+2}^l&=Z_{2i+1}Z_{2i+3}^{\red{\dagger}}\text{KT}_{l,N}.
 \end{align}
 \end{subequations}
 Hence $\text{KT}_{l,N}$ maps a nontrivial $\mathbb{Z}_N\times \mathbb{Z}_N$ SPT to two copies of  $\mathbb{Z}_N$ SSB phases.  
 \subsection{Two and higher dimensions}
In two dimensions SSPT phases protected by symmetry $G_s$ are classified by Ref.~\cite{devakul2018classification}
\begin{align}
    \mathcal{C}[G_s]\equiv H^2(G_s^2,U(1))/\left(H^2(G_s,U(1))\right)^3.
\end{align}
In our case $G_s=\mathbb{Z}_N\times\mathbb{Z}_N$ and the SSPT classification give $\mathcal{C}[\mathbb{Z}_N\times\mathbb{Z}_N]=\mathbb{Z}_N\times\mathbb{Z}_N\times \mathbb{Z}_N$. To enumerate various SSPT phases, we consider two square lattices with the color red and blue as in Fig.~\ref{fig:2drblattice}. The generators for the three $\mathbb{Z}_N$ factors are 1) $\mathbb{Z}_N\times\mathbb{Z}_N$ SSPT with cluster entangler between adjacent red and blue sites, 2) $\mathbb{Z}_N$ SSPT (generalization of $\mathbb{Z}_2$ SSPT we considered in two dimensions in section~\ref{sec:z2subsymmetricmodel}) on red lattice, and 3) $\mathbb{Z}_N$ SSPT on blue lattice. It should be straightforward to generalize the KT transformation to each of the three generators of the SSPT phase resulting in 1) two copies of the plaquette clock model, 2) one copy of the double-plaquette clock model, and 3) one copy of the double-plaquette clock model, respectively.

\section{The $\mathbb{Z}_2$ SSPT}
\label{sec:symmetrybreaking}
On a 2d square lattice, where each site contains 2 qubits, one can define the $\mathbb{Z}_2\times \mathbb{Z}_2$ SSPT for subsystem symmetries. The SSPT state is equivalent to the 2d cluster state given by the Hamiltonian 
\begin{widetext}
\begin{align}
\begin{split}
    \bm H_{\text{2dcluster}}&=-\sum_{v^r}X_{v^r}\prod_{v^b\in \partial p^b}Z_{v^b}-\sum_{v^b}X_{v^b}\prod_{v^r\in \partial p^r}Z_{v^r} -\lambda\sum_{v^r} X_{v^r}-\lambda\sum_{v^b} X_{v^b},
\end{split}
\end{align}
whereas each site now contains a blue (b) qubit and to its bottom left, a red (r) qubit, as illustrated in Fig.~\ref{fig:clustersquare}.

\begin{figure}[]
      \centering
      \includegraphics[width=0.3\textwidth]{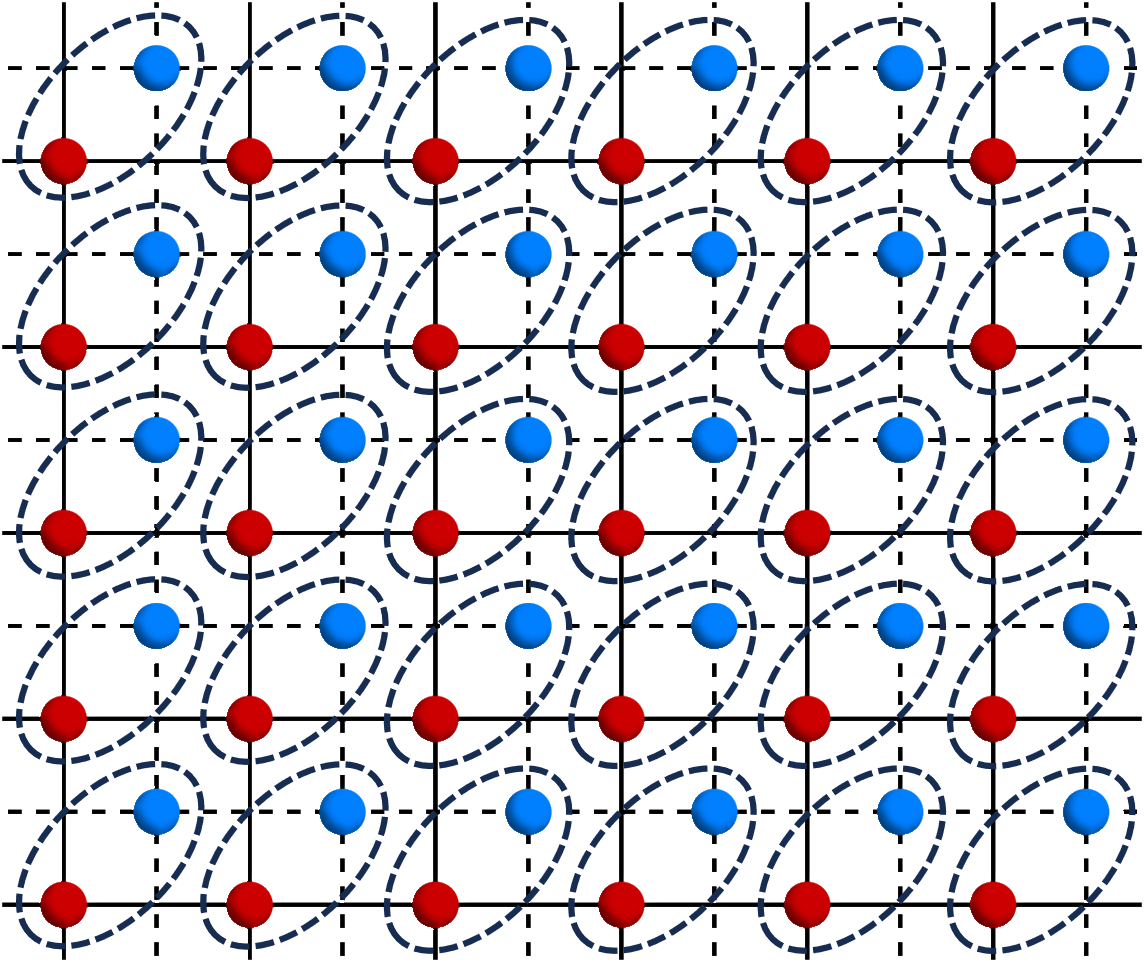}
      \caption{It can be regarded as a square lattice where each site contains a red qubit and a blue qubit.}
      \label{fig:clustersquare}
  \end{figure}

After inserting the symmetry-breaking terms, the Hamiltonian becomes
\beq
    \bm H_{\text{2dcluster}}=&-\sum_{v_r}X_{v^r}\prod_{v^b\in \partial p^b}Z_{v^b}-\sum_{v^b}X_{v^b}\prod_{v^r\in \partial p^r}Z_{v^r} -\lambda\sum_{v^r} X_{v^r}-\lambda\sum_{v^b} X_{v^b}-g\sum_{v^r}\begin{matrix}
        & Z_{v^b}\\
        Z_{v^r}&
    \end{matrix}.
\eeq
\end{widetext}
Tuning $g\rightarrow\infty$, the local Hilbert space of each site effectively reduces to two-dimensional in the low energy sector, since 
\beq
    \begin{matrix}
        & Z_{v^b}\\
        Z_{v^r}&
    \end{matrix}\equiv1.
\eeq
One can therefore map this reduced local Hilbert space into a one-qubit Hilbert space, i.e.,
\beq
    Z_{v^b},Z_{v^r}\rightarrow Z_v,\ \begin{matrix}
        & X_{v^b}\\
        X_{v^r}&
    \end{matrix}\rightarrow X_v.
\eeq

Under the perturbation theory, it can be shown that the low-energy effective Hamiltonian under the map becomes (after a re-scaling of energy)
\beq
    H=\frac{1}{2}\sum \begin{matrix}
        & Z& Z\\
        Z& X& Z\\
        Z& Z& \\
    \end{matrix}-\lambda\sum \begin{matrix}
        Z& Z\\
        Z& Z
    \end{matrix}-\frac{\lambda^2}{2}\sum X.
\eeq
For small $\lambda$, this model is in a $\mathbb{Z}_2$ SSPT order for the horizontal and vertical subsystem symmetries. 

In particular, when $\lambda=0$, we directly show here that the $\beta(g)$ phase defined in Ref.~\cite{devakul2018classification} is nontrivial. For an SSPT state, because of its short-ranged entangled nature, the truncated symmetry operator $U^{j_0,j_1}_{i_0,i_1}\equiv \prod_{i=i_0,j=j_0}^{i_1,j_1}X_{\red{i,j}}$ only affects locally on the corners,
\beq
    U^{j_0,j_1}_{i_0,i_1}V_{i_0,j_0}V_{i_1,j_0}V_{i_0,j_1}V_{i_1,j_1}\ket{\psi}=\ket{\psi},
\eeq
where operator $V_{i,j}$ is supported only locally around $(i,j)$. It can be easily shown that the operator 
\beq
V_{i,j}=Z_{i-1,j-1}Z_{i,j}.
\eeq
The characterizing phase for a nontrivial SSPT states is 
\beq
    \beta_{i,j}\equiv\bra{\psi}S^{\dagger}_i V_{i,j}^{\dagger}S_i V_{i,j}\ket{\psi}=-1,
\eeq
where $S_i\equiv\prod_{i'\geq i}\prod_{j}X_{i^{\red{'}},j}$ is the symmetry action on a half-plane. Therefore, the ground state when $\lambda=0$ is a strong SSPT protected by a linearly symmetric local unitary (LSLU). We note that a similar calculation can be performed for the 3D $\mathbb{Z}_2$ SSPT model.
\section{Measurement-based gauging method}
\label{app:Measurement-based}

In this section, we provide a measurement-based method to gauge the symmetries of various models that we consider in the main text in sections~\ref{sec:KWtwoandhigher} and \ref{sec:z2subsymmetricmodel}. 

\subsection{Plaquette Ising model}
\begin{figure}[h!]
    \centering
    \includegraphics[scale=1]{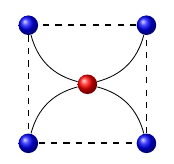}
    \caption{Cluster entanglement pattern for the KW duality map that maps between two plaquette Ising models. The blue lattice is the original lattice, and the red lattice is the ancilla lattice.  }
    \label{fig:plaquette_lattice}
\end{figure}

The plaquette Ising model has both horizontal and vertical subsystem symmetries. To gauge these subsystem symmetries, we introduce an ancilla d.o.f. in $\ket{+}$ state at each center of a plaquette $p \in \Delta_p$ (a vertex of the dual lattice). This d.o.f. is entangled to the vertices ($v \in \Delta_v$) of the corresponding plaquette via controlled-Z gate as shown in Fig.~\ref{fig:plaquette_lattice}. 
Then, we measure away the vertex d.o.f. in the Pauli X basis with the measurement outcome being all $\ket{+}$ state. 
Explicitly, this procedure can be written as an operator 
\begin{align}
    \text{KW}=\bra{+}^{\otimes\Delta_v}\prod_{p}\prod_{v\in\partial p}CZ_{p,v}\ket{+}^{\otimes\Delta_p}.
\end{align}
Then, KW satisfies 
\begin{align}
    \text{KW}X_v=\prod_{p,v\in\partial p}Z_p\text{KW},\\
    \text{KW}\prod_{v\in\partial p}Z_v=X_p\text{KW},
\end{align}
mapping between the plaquette Ising models on the original and dual lattices. This allows a short-depth implementation of the KW transformation and can be easily generalized to higher dimensions.
When measurement outcomes do not appear as the product state $|+ \rangle$, one can perform feedforward corrections to clean up non-trivial effects on the output wave function caused by them.
Such a mechanism was explained in Refs.~\cite{tantivasadakarn2021long, PhysRevB.108.115144} for gauging 0-form symmetries, and will be explained in Ref.~\cite{ops_2} for more general spin models. 

\subsection{Double plaquette Ising model}

\begin{figure}[h!]
    \centering
    \includegraphics[scale=0.8]{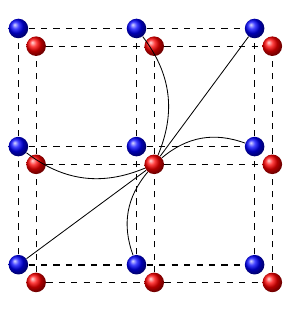}
    \caption{Cluster entanglement pattern for the KW duality map that maps between two double-plaquette Ising models. The blue lattice is the original lattice, and the red lattice is the ancilla lattice. }
    \label{fig:double_triangle_lattice}
\end{figure}

In addition to the horizontal and vertical symmetries, the double plaquette Ising model also has a diagonal symmetry. 
To implement the self-dual transformation of the double plaquette Ising model, we introduce another copy of the square lattice on top of the original lattice which we call the ancilla lattice. 
Let us denote the superimposed vertices on the two lattices by $v$ and $v^{(a)}$, where the superscript $(a)$ denotes the lattice on which the ancilla d.o.f. is initialized in the $\ket{+}$ state. 
A particular $v^{(a)}$ is entangled via controlled-Z gate to adjacent vertices on the original lattice as shown in Fig.~\ref{fig:double_triangle_lattice}. Let us denote the set of neighboring vertices (including diagonal but not off-diagonal) of an ancilla vertex $v^{(a)}$ by $\text{Neigh}(v^{(a)})$ and the set of neighboring ancilla vertices (including diagonal but not off-diagonal) of a vertex $v$ by $\text{Neigh}(v)$. 
Then, we measure away the vertex d.o.f. in the original lattice in the Pauli X basis with measurement outcome being all $\ket{+}$ state. 
Explicitly, this procedure is given by
\begin{align}
    \text{KW}=\bra{+}^{\otimes\Delta_v}\prod_{v^{(a)}}\prod_{v\in \text{Neigh}(v^{(a)})}CZ_{v^{(a)},v}\ket{+}^{\otimes\Delta_{v^{(a)}}}
\end{align}
KW satisfies
\begin{align}
    \text{KW}X_v=\prod_{v^{(a)}\in \text{Neigh}(v)}Z_{v^{(a)}}\text{KW},\\
    \text{KW}\prod_{v\in \text{Neigh}(v^{(a)})} Z_v= X_{v^{(a)}} \text{KW}.
\end{align}
mapping between the double plaquette Ising models on the original and ancilla lattices. 
This allows us to perform the KW duality and can be generalized to higher dimensions. Similar to the discussion in the plaquette Ising model, when measurement outcomes do not appear as the product state $\ket{+}$, one can perform the feedforward corrections to clean up the non-trivial effect on the output wavefunction caused by them \cite{ops_2}.
\subsection{Implementing the KT transformation}
We have seen in the previous subsections how to implement the KW transformations. Equipped with this, we can then use the duality web to perform the KT transformation as well by composing, e.g., $TST$, where $T$ can be implemented by the cluster-state entangling operation (and $S$ is the KW operation).

\red{
\section{Mapping of order parameters}\label{ref:order-param}
Order parameters of non-trivial phases get mapped under the KT transformations. Let $B_{k}$ be a product of $Z$ operators in the models we have considered in this work exhibiting SSSB phases, such as the plaquette Ising term, the double plaquette Ising term, etc. Let $K_k$ be the stabilizer of the corresponding SSPT cluster states. 
In SSSB phases, as a generalization of the two-point function $\langle Z_j Z_{j+\ell}\rangle$ in the 1d Ising model, a long-range entanglement can be detected by the order parameter 
\begin{align}
\langle \prod_{k \in \mathcal{M} } B_k \rangle,
\end{align}
where $\mathcal{M}$ is a segment consisting of cells (hypercubes) along a line where the subsystem symmetries are supported.
An example in the 2d plaquette Ising model is the product of $B_k = Z_{i,j} Z_{i+1,j} Z_{i+1,j} Z_{i+1,j+1}$ along a 1d segment (a thin strip consisting of plaquettes), which becomes a product of four Pauli $Z$'s at the corners of the strip: $\langle  Z_{i,j} Z_{i+\ell,j} Z_{i+\ell,j+1} Z_{i,j+1} \rangle$.
Upon the KT transformation, the corresponding SSPT string order parameter is written as 
\begin{align}
\langle \prod_{k \in \mathcal{M} } K_k \rangle, 
\end{align}
where the segment $\mathcal{M}$ is filled with $X$ operators in addition to the $Z$ operators at the corners.
For the 2d plaquette Ising model, an example of the string order parameter obtained by the KT operator~\eqref{eq:KT-2d-Z2Z2} is  $\langle  Z_{i,j} Z_{i,j+1}
X_{i+\frac{1}{2},j+\frac{1}{2}} \cdots X_{i+\ell - \frac{1}{2},j+\frac{1}{2} }
Z_{i+\ell,j} Z_{i+\ell,j+1} \rangle$.}

\end{document}